\documentclass[twocolumn]{aastex631}
\usepackage{amsmath}
\usepackage{multirow}

\usepackage{xcolor}

\shorttitle{Sensitivity of galaxies to warm dark matter}
\shortauthors{Costanza, Wang et al.}

\begin{document}

\title{On the sensitivity of different galaxy properties to warm dark matter}


\author[0009-0008-6301-1539]{Belén Costanza}
\affiliation{ Facultad de Ciencias Astron\'omicas y Geof\'isicas, Universidad Nacional de La Plata, Observatorio Astron\'omico, \\ Paseo del Bosque, B1900FWA La Plata, Argentina }
\affiliation{Consejo Nacional de Investigaciones Cient\'ificas y T\'ecnicas (CONICET), Rivadavia 1917, Buenos Aires, Argentina}

\author[0000-0001-7168-8517]{Bonny Y. Wang}
\affiliation{McWilliams Center for Cosmology and Astrophysics, Carnegie Mellon University, Pittsburgh, PA 15213, USA}
\affiliation{Entertainment Technology Center, Carnegie Mellon University, Pittsburgh, PA 15213, USA}

\author[0000-0002-4816-0455]{Francisco Villaescusa-Navarro}
\affiliation{Center for Computational Astrophysics, Flatiron Institute, 162 5th Avenue, New York, NY 10010, USA}
\affiliation{Department of Astrophysical Sciences, Princeton University, Peyton Hall, Princeton, NJ 08544, USA}

\author[0000-0002-8111-9884]{Alex M. Garcia}
\affil{University of Virginia 530 McCormick Rd Charlottesville, VA 22904, USA}
\affil{Virginia Institute for Theoretical Astronomy, University of Virginia, Charlottesville, VA 22904, USA}
\affil{The NSF-Simons AI Institute for Cosmic Origins, USA}

\author[0000-0002-2628-0237]{Jonah C. Rose}
\affiliation{Center for Computational Astrophysics, Flatiron Institute, 162 5th Avenue, New York, NY 10010, USA}
\affiliation{Department of Physics, Princeton University, Princeton, NJ 08544, USA}

\author[0000-0001-8593-7692]{Mark Vogelsberger}
\affil{Department of Physics, Kavli Institute for Astrophysics and Space Research, Massachusetts Institute of Technology, Cambridge, MA 02139, USA}
\affil{The NSF AI Institute for Artificial Intelligence and Fundamental Interactions, Massachusetts Institute of Technology, Cambridge, MA 02139, USA}

\author[0000-0002-5653-0786]{Paul Torrey}
\affiliation{Department of Astronomy, University of Virginia, Charlottesville, VA 22904, USA}
\affiliation{Virginia Institute for Theoretical Astronomy, University of Virginia, Charlottesville, VA 22904, USA}
\affiliation{The NSF-Simons AI Institute for Cosmic Origins, USA}

\author[0000-0003-0777-4618]{Arya~Farahi}
\affiliation{Departments of Statistics and Data Sciences, University of Texas at Austin, Austin, TX 78757, USA}
\affiliation{The NSF-Simons AI Institute for Cosmic Origins, USA}

\author[0000-0002-6196-823X]{Xuejian Shen}
\affiliation{Department of Physics, Kavli Institute for Astrophysics and Space Research, Massachusetts Institute of Technology, Cambridge, MA 02139, USA}

\author[0009-0006-5740-5318]{Ilem Leisher}
\affiliation{Grinnell College, 1115 8th Ave, Grinnell, IA 50112, USA}
\affiliation{Department of Physics, Massachusetts Institute of Technology, Cambridge, MA 02139, USA}
\affiliation{Kavli Institute for Astrophysics and Space Research, Massachusetts Institute of Technology, Cambridge, MA 02139, USA}

\begin{abstract}

We study the impact of warm dark matter (WDM) particle mass on galaxy properties using 1,024 state-of-the-art cosmological hydrodynamical simulations from the DREAMS project. We begin by using a Multilayer Perceptron (MLP) coupled with a normalizing flow to explore global statistical descriptors of galaxy populations, such as the mean, standard deviation, and histograms of 14 galaxy properties. We find that subhalo gas mass is the most informative feature for constraining the WDM mass, achieving a determination coefficient of $R^2 = 0.9$. We employ symbolic regression to extract simple, interpretable relations with the WDM particle mass.
Finally, we adopt a more localized approach by selecting individual dark matter halos and using a Graph Neural Network (GNN) with a normalizing flow to infer the WDM mass, incorporating subhalo properties as node features and global simulation statistics as graph-level features. The GNN approach yields only a residual improvement over MLP models based solely on global features, indicating that most of the predictive power resides in the global descriptors, with only marginal gains from halo-level information.

\end{abstract}


\section{Introduction} \label{sec:intro}

Dark matter (DM) is a fundamental component of the Universe whose existence is needed to explain many different cosmological observations. Specifically, the Cold Dark Matter (CDM) paradigm, which assumes that dark matter behaves as a collisionless and pressureless fluid, has been highly successful in describing the large-scale structure of the Universe~\citep{2005Natur.435..629S}. The CDM model predicts that the structure grows hierarchically through gravitational instability, with cold, low-velocity DM particles clustering efficiently~\citep{1984Natur.311..517B, 1985ApJ...292..371D}.

The CDM model is not without its faults; several discrepancies between the CDM model and observational data on galactic and subgalactic scales have emerged. Some of the well-known tensions are: the abundance of dwarf galaxies called ``missing satellites (MS) problem" for galaxies in the Milky Way~\citep[e.g.][]{1999ApJ...524L..19M, 1999ApJ...522...82K}, the ``too-big-to-fail (TBTF) problem"~\citep{2011MNRAS.415L..40B, 2014MNRAS.444..222G, 10.1093/mnras/stu474, 2015A&A...574A.113P}, the ``core-cusp" (CC) problem~\citep[e.g.][]{1997MNRAS.290..533D}, among others.

Some of these tensions can be alleviated by considering the effects of baryonic processes in galaxy formation models. For instance, the ``core-cusp" issue may be mitigated by sufficiently bursty supernova (SN) feedback~\citep[e.g.][]{2020MNRAS.497.2393L}, while the ``missing satellites" problem can be addressed by considering gas heating effects during the reionization epoch~\citep{2014Natur.509..177V, 2015MNRAS.454.2981C}. However, it remains unclear whether these discrepancies demand modifications to the treatment of galaxy formation physics or suggest the need for modifications to the dark matter nature itself. In particular, other tensions, such as the TBTF problem, may require alternative DM models~\citep{10.1093/mnras/sty2339}. 


Since no direct detection of DM particles has yet provided insights into its nature, it is feasible to consider DM physics beyond the CDM model. Several alternatives exist as small modifications to the CDM model; for example, the Warm Dark Matter (WDM) model~\citep{PhysRevD.62.063511, Sommer-Larsen_2001} addresses some of the galaxy formation challenges while preserving the success of the CDM framework at large cosmological scales. WDM particles possess appreciable thermal velocities, enabling them to escape from shallow potential wells in the early Universe. This characteristic impacts small-scale structure formation, resulting in the suppression of the abundance of small halos and a reduction in the central densities of halos~\citep{2001ApJ...556...93B}.


Hierarchical structure formation is a highly non-linear process, making it challenging to solve analytically. As a result, this regime is primarily accessible through numerical cosmological simulations. In particular, projects like CAMELS~\citep{2023ApJS..265...54V} and DREAMS~\citep{2025ApJ...982...68R} provide extensive suites of simulations with systematic variations to cosmology and baryonic physics that can be used to extract cosmological information or constrain DM models while accounting for uncertainties in baryonic physics. 

One of the key strengths of the DREAMS suite lies in its potential to provide robust constraints on the WDM particle mass by considering a statistically significant sample of galaxies. Current limits on the WDM were derived from relatively small samples, such as the Milky Way satellite population or strong lensing systems~\citep{2021ApJ...917....7N, 2021PhRvL.126i1101N}.

To distinguish between DM models several statistics have been studied, such as the power spectrum and the abundance of subhalos. In this context, machine learning techniques have emerged as powerful tools to extract maximal information from cosmological fields, enabling precise constraints on cosmological parameters. Recent studies include the application of Convolutional Neural Networks (CNNs) to marginalize over baryonic processes and predict the underlying cosmology~\citep{Villaescusa-Navarro_2021}, the inference of WDM particle masses with CNNs~\citep{2024MNRAS.527..739R}, and the identification of tight correlations between the total matter density of the universe $\Omega_{m}$ and individual galaxy properties~\citep{2022ApJ...929..132V, Echeverri-Rojas_2023, 2024ApJ...969..105C, Wang2024}, among others. 

In a recent study, \cite{2024ApJ...970..170L} studied the feasibility of constraining the masses of WDM particles using individual galaxy properties, finding no significant correlation between them. In this work, we investigate whether statistical properties of galaxy populations or subhalo systems within individual halos contain more robust information on the nature of dark matter.

We first consider the distribution of galaxy properties in simulations (e.g. the abundance of subhalo masses) as the statistics and quantify how accurately we can infer the WDM mass. To this end, we train a Multilayer Perceptron model (MLP) combined with Normalizing Flows (NFs) to learn the posterior distribution $p(\vec{\theta}|\textbf{X})$, where $\vec{\theta}$ represents the WDM mass parameter, and \textbf{X} corresponds to summary statistics of the galaxy population. We complement this with symbolic regression~\citep{Wilstrup2021, Reinbold2021, LaCava2021, pysr} techniques to provide interpretability for the features with the most predictive power.  

Next, we focus our attention on galaxies in individual halos and quantify whether they can be used to infer the WDM mass. We employ Graph Neural Networks (GNNs) to process subhalo populations of single halos, again using NFs to model the posterior over $\vec{\theta}$. GNNs have shown strong performance in inferring cosmological parameters~\citep{2022ApJ...937..115V} and halo masses~\citep{2022ApJ...935...30V} from point clouds. NFs have also been successfully applied to infer cosmological parameters given galaxy photometry~\citep{2023mla..confE..14H}, and have been incorporated into deep generative models to reconstruct halo assembly histories~\citep{2024MNRAS.533.3144N}, among other applications.

This paper is organized as follows. In Sec.~\ref{sec:data} we describe the hydrodynamical simulations and galaxy properties used to construct the training, validation, and testing datasets. In Sec.~\ref{sec:GNN} we outline the architecture of the MLP + NF and GNN + NF models, and also describe the symbolic regression framework. Subsequently, in Sec.~\ref{sec:methods} we detail the methods employed for the training of the models. We then present the results of the WDM inference using both statistical and individual-halo approaches in Sec.~\ref{sec:results}. We conclude in Sec.~\ref{sec:conclusions} by discussing our findings and their implications.

\section{Dataset} \label{sec:data}

In this work, we use cosmological hydrodynamic simulations from the \textsc{DREAMS} (DaRk MattEr and Astrophysics with Machine learning and Simulations) project\footnote{https://www.dreams-project.org/} \citep{2025ApJ...982...68R}. DREAMS contains thousands of simulations that are organized based on the dark matter model adopted (e.g., cold dark matter vs. alternative dark matter), the environment simulated (full cosmological boxes vs. targeted zoom-ins), and the employed galaxy formation model. The goal of the project is to identify distinct signatures of non-CDM models that remain robust despite astrophysical uncertainties inherent to galaxy formation modeling. To achieve this, DREAMS provides simulation suites designed specifically for training machine learning models capable of isolating the unique effects of modified dark matter scenarios, while marginalizing over a wide range of uncertain astrophysical parameters.



Specifically, we utilize cosmological boxes that track the evolution of $256^{3}$ dark matter particles and $256^{3}$ initial gas resolution elements within a periodic comoving volume of $(25h^{-1} \rm{Mpc})^{3}$ from $z=127$ down to $z=0$. The DM mass resolution of the simulations is $7.81 \times (\Omega_{\rm m}/0.302) \times 10^{7}~h^{-1} M_{\odot}$ and the baryon mass resolution is $1.27 \times 10^{7}~h^{-1} M_{\odot}$. The gravitational softening is equal to $1.0~h^{-1} \rm{kpc}$ at redshift $z=0$.

Uniform boxes have the advantage of providing large statistics by covering a significant volume with a large number of galaxies, but at the cost of lower resolution. By utilizing information from thousands of galaxies, we can explore how DM models (specifically WDM models) influence the statistical properties of galaxies. We note, however, that these simulations do not model galaxies with a stellar mass below $~10^{7} M_{\odot}$~\citep{2023MNRAS.522.3831F, 2019MNRAS.490.3234N}, which prevents us from distinguishing deviations from cold dark matter models on small scales. 


The simulations used in this work were run with the \textsc{Arepo} code~\citep{2010MNRAS.401..791S, 2020ApJS..248...32W} and were executed using the IllustrisTNG galaxy formation model~\citep{2017MNRAS.465.3291W, 2018MNRAS.473.4077P}, which itself is based on the Illustris model~\citep{2013MNRAS.436.3031V, 2014MNRAS.438.1985T}. The specific simulation suite we employ consists of 1,024 independent simulations, all based on the fiducial TNG model setup, but spanning a range of cosmological and astrophysical parameters, as well as initial random seeds.


All simulations share the values of these cosmological parameters: $\Omega_{\rm b} = 0.049$, $h=0.6711$, $n_{s} = 0.9691$, $w = -1$, $M_{\nu} = 0.0eV$, $\Omega_{\rm k} = 0.0$. However, the values of $\Omega_{\rm m}$,  $\sigma_8$, and three astrophysical parameters ($A_{\rm {SN1}}$, $A_{\rm{SN2}}$, $A_{\rm{AGN}}$) are different in each simulation. The varied astrophysical parameters control the efficiency of supernova and active galactic nuclei feedback, respectively. 
The values of these parameters are arranged in a Sobol sequence \citep{sobo:1967:russ} with ranges: 

\begin{eqnarray}
    0.1 \leq \Omega_{\rm m} \leq 0.5, \\
    0.6 \leq \sigma_{8} \leq 1.0,  \\
    0.25 \leq A_{\rm{SN1}} \leq 4.0,  \\
    0.50 \leq A_{\rm{SN2}} \leq 2.0, \\
    0.25 \leq A_{\rm{AGN}} \leq 4.0, \\
    1.8 \ \rm{keV} \leq m_{\rm{WDM}} \leq 16 \ \rm{keV}. 
\end{eqnarray}
The values of $\Omega_{\rm m}$ and $\sigma_{8}$ are sampled linearly, while the values of the astrophysical parameters are sampled logarithmically. The choice of wide priors for $\Omega_{\rm m}$ and $\sigma_{8}$ is based on the work of~\cite{2025ApJ...982...68R}, where the ranges were deliberately set broader than the current uncertainties from CMB analysis~\citep{2020A&A...641A...6P}. This strategy minimizes the impact of the prior on the machine learning training process and ensures that our results are not artificially biased by overly restrictive assumptions.

The free parameters $A_{\rm{SN1}}$ and $A_{\rm{SN2}}$ serve as normalization factors that account for the energy injection rate and wind speed, respectively, of the galactic winds induced by supernova feedback, modeled as in \citep{10.1046/j.1365-8711.2003.06206.x}. The AGN parameter $A_{\rm{AGN}}$ governs the normalization factor of the AGN feedback in the high-accretion state. When the value of the astrophysical parameters is equal to 1, we recover the fiducial IllustrisTNG model.

\begin{figure}[h!]
\centering
\includegraphics[width=1\columnwidth]{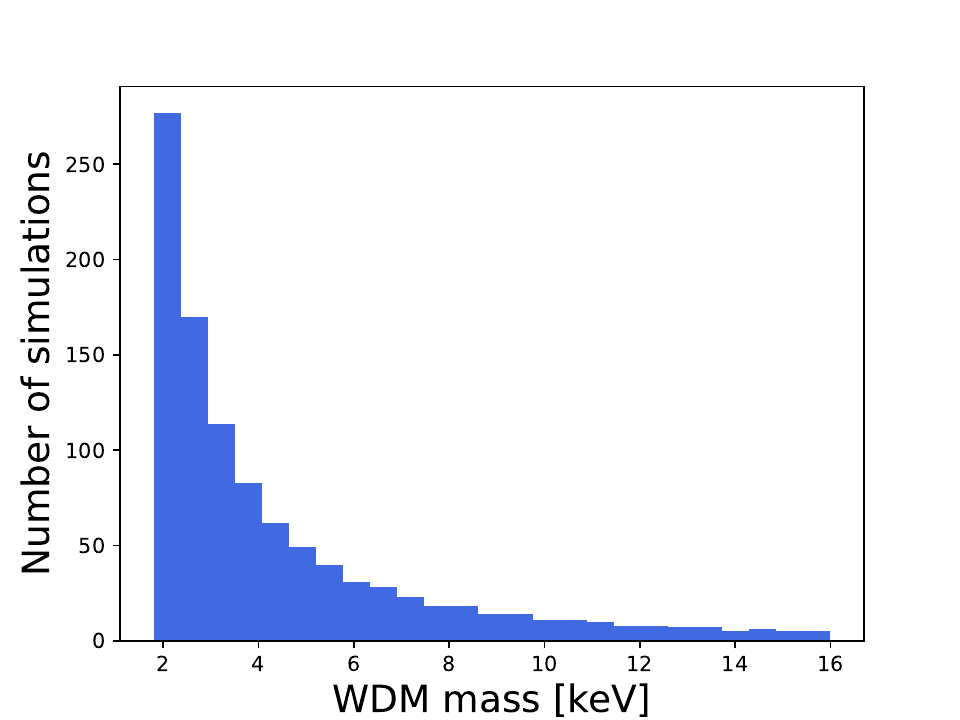}
\caption{Histogram of 1024 unique WDM particle masses used in the varied cosmology simulation suite, ranging between 1.8 keV and 16 keV, sampled uniformly from an inverse distribution.}
\label{fig:wdm_range}
\end{figure}

In addition, the WDM particle mass is different for each simulation, and it is sampled uniformly from an inverse distribution of particle masses. Figure~\ref{fig:wdm_range} presents the histogram of the WDM particle masses in the simulations, ranging from 1.8 keV to 16 keV. It can be observed that there are fewer simulations at higher WDM particle masses, which may limit the performance of the inference for these masses.

The WDM model adopted in our simulations assumes that warm dark matter particles are produced thermally. As a consequence, the linear matter power spectrum is suppressed on small scales relative to CDM, following $\rm{P_{wdm}} = \beta(k)^{2} \rm{P_{CDM}}$ where $\beta(k)$ is the transfer function that depends of the comoving wavenumber $k$. We refer the reader to~\cite{2025ApJ...982...68R} for details of the specific transfer function implemented in our simulations, and to~\citep{Bode_2001} where the WDM model was introduced.


The halos and subhalos in the simulations are identified using the friends-of-friends (FoF) and \textsc{subfind} algorithms~\citep{10.1046/j.1365-8711.2001.04912.x, 2009MNRAS.399..497D}. The FoF algorithm identifies halos based on the proximity of dark matter particles, while \textsc{subfind} refines these groups by detecting gravitationally bound substructures. By construction, \textsc{subfind} only outputs subhalos containing at least 20 bound resolution elements (sum over all particle types)\footnote{https://www.tng-project.org/data/}, establishing the effective resolution floor of the sample~\citep{2012MNRAS.423.1200O}. In the main analysis, we include all identified subhalos, regardless of whether they contain stars, since small subhalos are particularly relevant to our analysis. However, in Appendix~\ref{apx2}, we test the robustness of our results by applying two resolution thresholds: first restricting the sample to subhalos with more than 30 DM particles to ensure completeness, and then to subhalos with more than 50 DM particles for a more conservative selection.

For each subhalo, we consider the following 14 properties from the \textsc{subfind} catalog:

\begin{itemize}
    \item $\mathbf{M_{g}}$: the gas mass content. 
    \item $\mathbf{M_{BH}}$: the black-hole mass. 
    \item $\mathbf{M_{*}}$: the stellar mass. 
    \item $\mathbf{M_{t}}$: the total mass. 
    \item $\mathbf{V_{max}}$: the maximum circular velocity. 
    \item $\boldsymbol{\sigma_{\nu}}$: the velocity dispersion of all particles contained in the subhalo.
    \item $\mathbf{Z_{g}}$: the mass-weighted gas metallicity. 
    \item $\mathbf{Z_{*}}$: the mass-weighted stellar metallicity. 
    \item \textbf{SFR}: the subhalo star-formation rate.
    \item \textbf{J}: the modulus of the subhalo spin vector.
    \item \textbf{V}: the modulus of the subhalo peculiar velocity.
    \item $\mathbf{R_{*}}$: the radius containing half of the subhalo stellar mass.
    \item $\mathbf{R_{t}}$: the radius containing half of the total mass.
    \item $\mathbf{R_{max}}$: the radius at which $\sqrt{GM(\langle R_{\mathrm{max}})/R_{\mathrm{max}}}=V_{\mathrm{max}}$
\end{itemize}

These subhalo properties are used throughout our analysis. At the simulation level, we compute summary statistics, such as the mean, standard deviation, and the full distribution in the form of histograms, over all subhalos in a given simulation using these properties. At the halo level, the same features of the subhalos within each halo serve as inputs to our GNN, enabling a more detailed analysis of dark matter model effects on galaxy-scale structures.

\section{Models} 
\label{sec:GNN}

In this section, we present the architectures employed to infer the WDM particle mass from cosmological simulations, including multilayer perceptrons, graph neural networks, and normalizing flows. In addition, we briefly describe how symbolic regression works.

\subsection{Multilayer perceptron}

We employ multilayer perceptron layers to infer WDM masses from the statistical properties of galaxy features.

A Multilayer Perceptron (MLP)~\citep{1986Natur.323..533R} is a type of feedforward neural network composed of multiple layers of fully connected nodes. Each layer applies a linear transformation to its inputs, followed by a nonlinear activation function, allowing the network to learn complex, nonlinear relationships between inputs and outputs. In our case, the MLP receives as input a set of global statistical descriptors derived from the galaxy population in each simulation, such as the mean, standard deviation, or histogram bins of physical properties (detailed in Sec.~\ref{sec:data}). The MLP outputs a latent representation which is then used as input to a normalizing flow to model the posterior distribution of the WDM mass. This approach allows us to quantify the sensitivity of the statistical properties of galaxy features to the WDM mass.

\subsection{Graph neural networks}
\label{sec:graphdata}

We employ graph neural networks when inferring WDM masses from the subhalos of individual halos.



Graph Neural Networks (GNNs) are especially well suited for handling data characterized by arbitrary relational structures among entities~\citep{2018arXiv180601261B, 2021arXiv210413478B}. These entities are referred to as \textit{nodes}, and their relations are specified by the so-called \textit{edges}. Therefore, a \textit{graph} can be defined as a 3-tuple $G = (\textbf{u}, V, E)$, where \textbf{u} represents the global system-level properties, \textit{V} is the set of nodes, and \textit{E} is the set of edges, which represents the connectivity (directed or undirected) of the graph.

In our work, we consider halos that host multiple subhalos, and construct a graph where the subhalos serve as the nodes. Additionally, we define an undirected edge between two subhalos if their separation is smaller than a given threshold, specified by a hyperparameter known as the linking radius $r_{\rm{link}}$.

The set of nodes, denoted as $V = \{\mathbf{v}_{i}\}$, comprises node attributes $\mathbf{v}_{i}$ representing the 14 subhalo properties detailed in Sec.~\ref{sec:data}. The connectivity between two nodes $i$ and $j$ depends on the positions of the subhalos and is denoted by an edge $\mathbf{e}_{i,j}$. The positions of the subhalo nodes transform, under translations and rotations, as $\mathbf{p}_{i} = \mathbf{R}\mathbf{p}_{i}+\mathbf{T}$, where \textbf{R} and \textbf{T} denote the rotation and translation matrices, respectively. We build our input graph features so that they are invariant under the group E(3). 

To ensure translational symmetry, the edge features can be defined as the relative positions between nodes, i.e., $\mathbf{d}_{i,j} = \mathbf{p}_{i} - \mathbf{p}_{j}$. To account for rotational symmetry, we use scalar products of distances, denoted as $\alpha_{i,j} = \mathbf{n}_{i} \cdot \mathbf{n}_{j}$ and $\beta_{i,j} = \mathbf{n}_{i} \cdot \mathbf{s}_{i,j}$, where $\mathbf{n}_{i} = (\mathbf{p}_{i}-\bar{\mathbf{p}})/|\mathbf{p}_{i}-\bar{\mathbf{p}}|$, $\bar{\mathbf{p}}$ is the centroid of the distribution, and $s_{i,j} = \mathbf{d}_{i,j}/|\mathbf{d}_{i,j}|$. Then, the edge features connecting nodes are: 
\begin{equation}
    \mathbf{e}_{i,j} = [|\mathbf{d}_{i,j}|/r_{\rm{link}}, \alpha_{i,j}, \beta_{i,j}],
\end{equation}
where $r_{\rm{link}}$ is the linking radius. With this definition, the edge features $\mathbf{e}_{i,j}$ are both translational and rotational invariant. For more details about the edge features and symmetries, we refer the reader to \cite{2022ApJ...937..115V} who first proposed this scheme. 

So far, we have described the components of the graphs, which encode the subhalo information within individual halos. In addition to the node and edge features, each graph also includes global features, denoted by \textbf{u}, which represent properties of the entire halo. To compare the performance of the inference when incorporating halo-level information with the case using only global statistics (mean and standard deviation) from the cosmological box, we have used those statistics as the global features of the graph. Besides, we also study the improvement of the results adding $\Omega_{m}$ as a global feature.

\begin{figure}
\centering
\includegraphics[width=1\columnwidth]{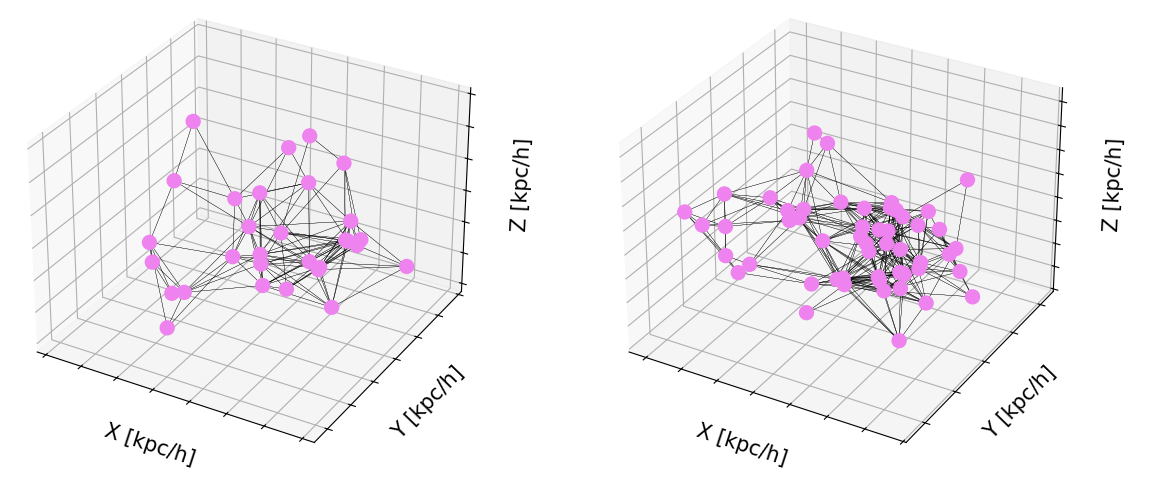}
\caption{Two graphs examples built from galaxy catalogs (two different DREAMS-IllustrisTNG simulations). The purple nodes represent the galaxies of the halo and they are connected if their distance is smaller than the linking radius $r_{\rm{link}}$.}
\label{fig:graph_example}
\end{figure}

To train the GNN, we constructed training and validation datasets, each consisting of graphs employed as inputs in the model. Figure \ref{fig:graph_example} presents two graphs examples from two different DREAMS-IllustrisTNG simulations showing the nodes and the edges connecting them.

The architecture of a graph neural network comprises multiple graph layers. Each layer takes a graph as input and returns the same graph topology as output, with updated features for its nodes and edges. As a first step, we built a GNN with the objective of creating a latent space representation. For that purpose, we follow the architecture of \textsc{CosmoGraphNet}~\citep{2022ApJ...937..115V} using the library \emph{PyTorch Geometric}~\citep{2019arXiv190302428F}. 

The $l$ graph layer updates the values of the edge features and node features. An edge model takes as input the edge feature and the two connected nodes, and it returns updated edge attributes:

\begin{equation}
    \mathbf{e'}_{i,j}  = g(\mathbf{v}_{i}, \mathbf{v}_{j}, \mathbf{e}_{i,j}).
\end{equation}

A node model takes as input the node features, global features, and the \textit{aggregation} of edge features: 

\begin{equation}
    \mathbf{v'}_{i} = f\left(\mathbf{v}_{i}, \bigoplus_{k\in N_{i}} \mathbf{e'}_{ki}, \mathbf{u}\right),  
\label{eq:node}
\end{equation}
where $g$ and $f$ are non-linear functions represented as multilayer perceptrons, and $N_{i}$ are all the nodes connected to the node $\mathbf{v}_{i}$. The operator $\bigoplus$ represents a permutation-invariant operation on the edges, aggregating information from the neighborhoods to the node $\mathbf{v}_{i}$. Examples of such operations include the mean, maximum, and sum, where in our model we consider a concatenation of them into a single vector. 

There are subsets of GNNs called DeepSets~\citep{2017arXiv170306114Z}, where the neighbors of a node are not taken into account (or in general do not exist) to update the node features. Instead, only the attributes of the nodes themselves are used to propagate information to the next layer. In this case, equation~\eqref{eq:node} becomes:
\begin{equation}
    \mathbf{v'}_{i} = f(\mathbf{v}_{i}, \mathbf{u}),  
\end{equation}
where the attributes of a node in a given layer depend on the values of the same node in the previous layer, plus the global features. In Sec.~\ref{sec:results}, we present the results considering both GNNs and DeepSets. The idea of using DeepSets over GNNs is to quantify how much information edges contain, or, in our case, how much information is embedded into spatial clustering.

Finally, after $K$ graph layers, there is a global \textit{pooling} layer that aggregates the information from all the updated nodes and the global features. Subsequently, a multilayer perceptron is applied, and the output of the neural network, denoted as \textbf{y}, is obtained:

\begin{equation}
    \mathbf{y} = h\left(\bigoplus_{i\in G}\mathbf{v}_{i}^{k}, \mathbf{u}\right).
    \label{yend}
\end{equation}

This output can represent a latent space that serves as input for the normalizing flow, or it can also be the predicted WDM mass, depending on the problem to be solved (inference or regression).

\subsection{Normalizing flows}
\label{sec:NF}

We employ normalizing flows to learn probability distribution functions $p(\vec{\theta}|\mathbf{X})$, where $\vec{\theta}$ is the parameter of interest (WDM mass in our case), and $\mathbf{X}$ is a summary statistics (for instance, the latent space of a GNN or MLP).





 
Normalizing flows (NFs)~\citep{Jimenez2015} employ neural networks to learn a highly flexible and bijective transformation \( f : \mathbf{z} \mapsto \mathbf{x} \), which maps a simple base distribution \( \pi(\mathbf{z}) \) into a more complex target distribution over \textbf{x}. The function \( f \) is designed to be invertible and have a manageable Jacobian determinant, allowing the target distribution to be obtained from \( \pi(\mathbf{z}) \) via the change of variables formula. Since \( \pi(\mathbf{z}) \) is easy to evaluate, this also enables efficient computation of the target density:

\begin{equation}
    p(\mathbf{x}) = \pi\left(f^{-1}_\phi(\mathbf{x})\right)
    \left|
    \det\left(\frac{\partial f^{-1}_\phi}{\partial \mathbf{x}}\right)
    \right|.
\end{equation}
In our context, the target distribution is the posterior distribution of the warm dark matter particle mass, conditioned on a latent representation extracted from either the MLP or the GNN, while the base distribution is a simple Gaussian. 

The overall transformation function $f_{\phi}$ is constructed by composing several invertible transformations, each parameterized by a neural network with parameters $\phi$. Specifically, we employ neural spline flows~\citep{2019arXiv190604032D, 2020arXiv200105168D}, where each transformation is defined using a spline function. These splines form a family of smooth, piecewise functions composed of polynomial segments, each acting over different intervals of the domain.

We implement spline autoregressive flows, in which each transformation is applied in an autoregressive manner. In this setting, the target distribution is factorized as: 

\begin{equation}
    p(\mathbf{x}) = \prod_{i}^{D} p(x_{i}|x_{1;i-1}), 
\end{equation}
where $x_{1;i-1} = [x_{1},x_{2},...,x_{i-1}]$ and $D$ is the dimensionality of the target distribution. Each conditional distribution $p(x_{i}|x_{1;i-1})$ is modeled with a spline function that depends on the previous dimensions $x_{<i}$. The parameters of the spline are optimized through autoregressive layers. 

The hyperparameters of the NF model include the number of conditional layers (i.e., the number of transformations within the flow), the number of bins (spline segments), and the number of hidden dimensions in the autoregressive neural network. We have used the \emph{Pyro} package~\citep{bingham2019pyro} to define and train the model.
Finally, the loss function to be minimized is:
\begin{equation}
    \mathcal{L} = -\mathrm{log}[p(\vec{\theta}|\mathbf{X})].
\label{eq:loss}
\end{equation}

\begin{figure*}
    \centering
    \includegraphics[width = 0.48\linewidth]{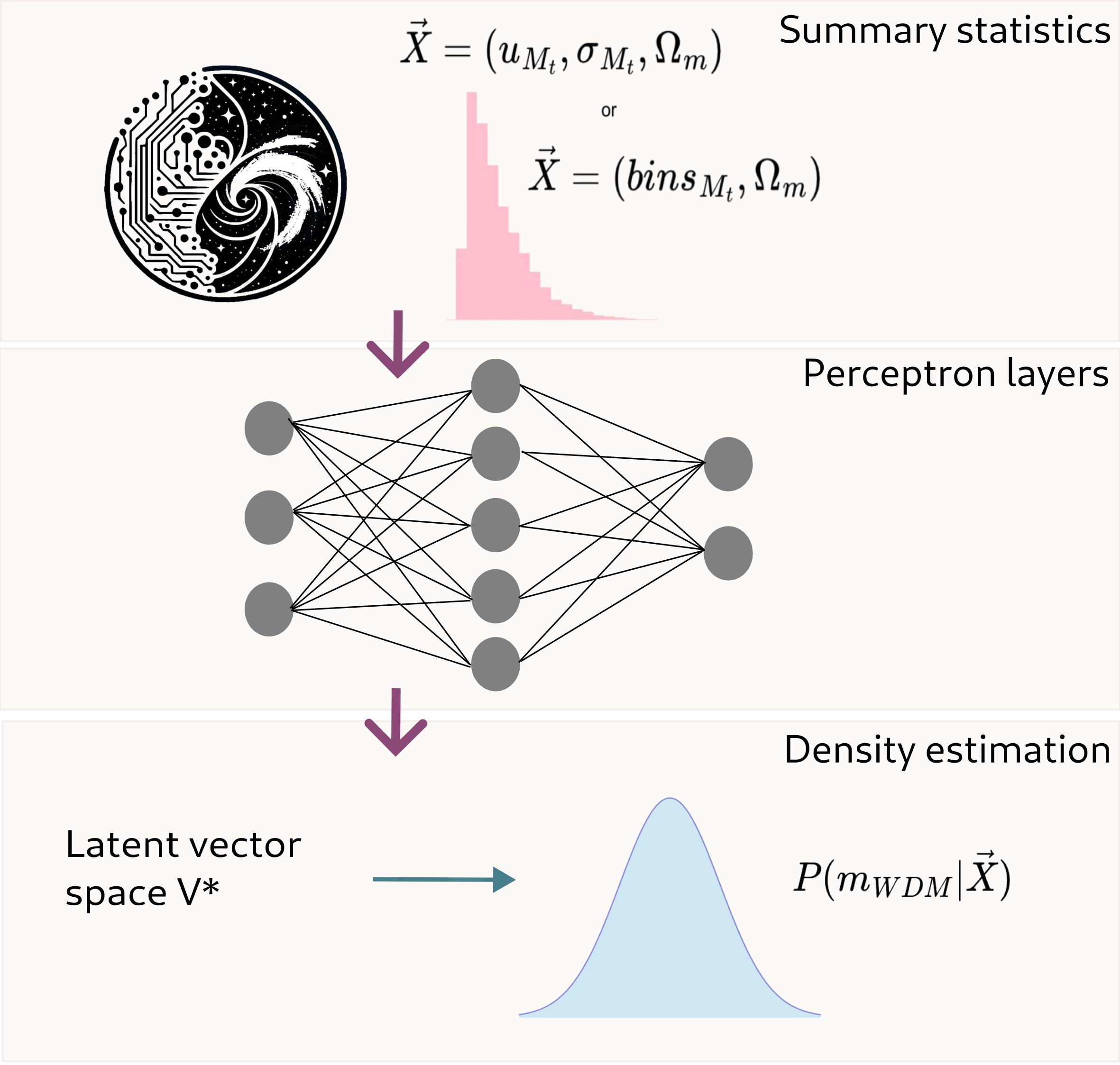}
    \includegraphics[width = 0.48\linewidth]{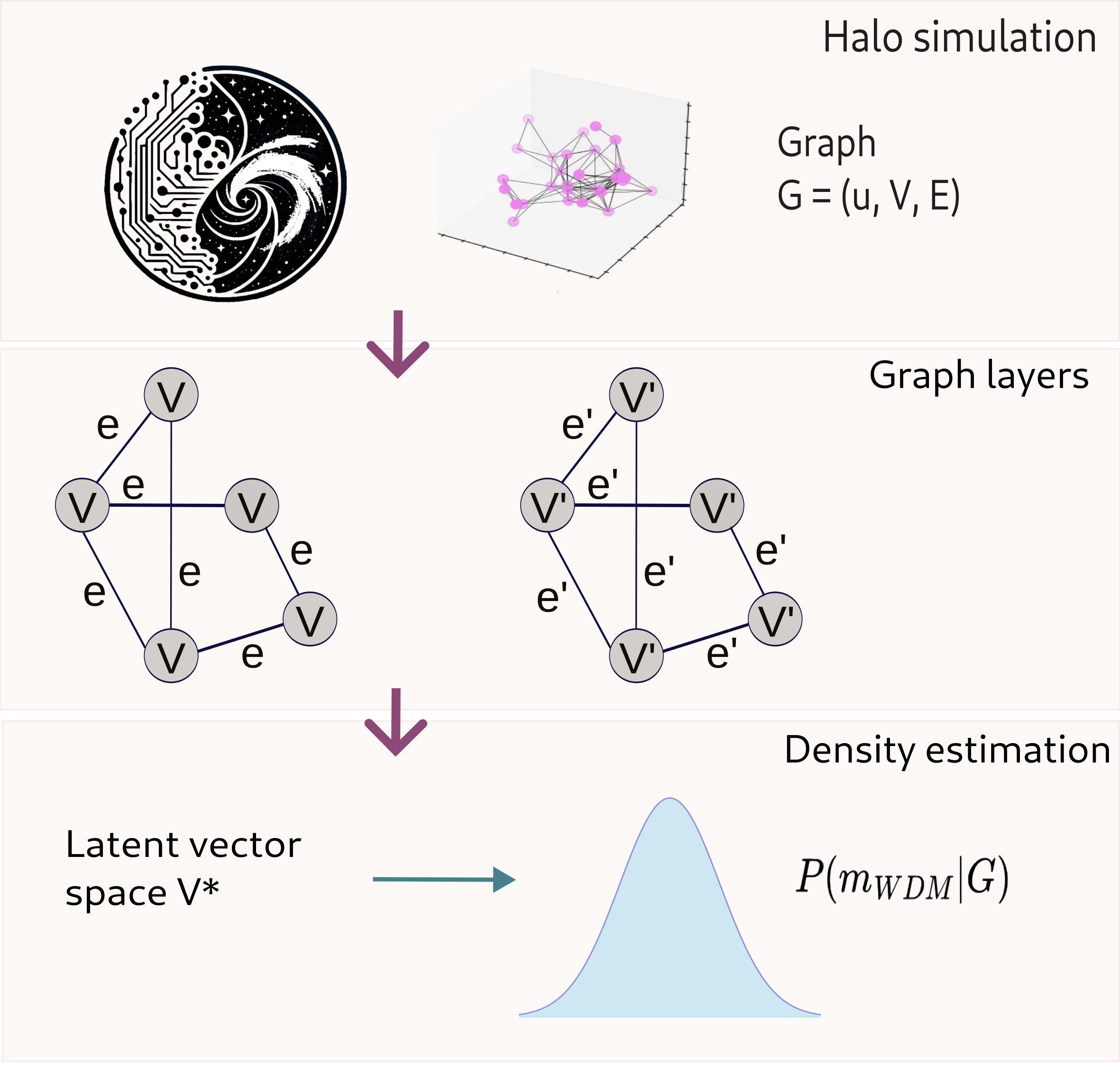}
    \caption{Architectures used for inferring the WDM particle mass from the DREAMS simulations. Left panel: the input consists of global statistical descriptors of galaxy properties across the simulation box, such as the mean, standard deviation, or histogram bins. These features pass through multiple perceptron layers (two are shown for illustration), producing a latent representation that serves as input to the normalizing flow. Right panel: the graphs are constructed from the subhalos hosted by the halos. These graphs pass through multiple graph layers (two are shown), which produce a latent space representation. In both cases, the resulting latent vector is passed to the normalizing flow, which outputs the posterior distribution of the WDM mass conditioned on either the global descriptors or the halo-specific graph.}
\label{fig:draw}
\end{figure*}

Figure~\ref{fig:draw} illustrates the architecture of the models employed in this work. The left panel shows the MLP + NF architecture, where the input consists of global statistical descriptors of galaxy properties. These features are passed through several perceptron layers to produce a latent representation. The right panel presents the GNN + NF architecture, where the input is a graph constructed from the subhalos within a single dark matter halo. This graph is processed through multiple graph layers to generate its own latent representation. In both cases, the resulting latent space is fed into a normalizing flow, which outputs the posterior distribution of the WDM particle mass conditioned on either the global descriptors or the halo-specific graph structure.

\subsection{Symbolic Regression}

In addition to the models described above, we use symbolic regression to obtain more interpretable mathematical equations for our analysis. The compactness of equations reduces the computational cost of the models and enhances their ability to accurately recover the underlying mathematical structure of a given dataset. Moreover, symbolic regression tends to produce formulas that are more robust to noise than many traditional machine learning approaches \citep{Wilstrup2021, Reinbold2021, LaCava2021}. They often demonstrate superior generalization performance overall. Recently, this technique has been successfully developed and applied to a variety of problems in astrophysics (e.g., \citealt{Cranmer2020}; \citealt{Wong2022}; \citealt{Shao2023}; \citealt{Wadekar2023}; \citealt{Lemos2023}).

Specifically, in this paper, we use the symbolic regression model from \emph{PySR} \citep{pysr}. \emph{PySR} employs evolutionary algorithms, which are techniques inspired by biological evolution and natural selection, to strategically search the space of mathematical models. These algorithms enable efficient exploration of potential expressions, offering a favorable balance between predictive accuracy and computational complexity. In particular, we apply symbolic regression to extract interpretable equations that relate the warm dark matter mass to the most important summary statistics, identified as key global features by our machine learning models. This approach provides both physical insight and practical formulas that connect galaxy properties to the underlying warm dark matter mass.

\section{Methods} 
\label{sec:methods}

The trained models, using different input data, can be summarized as follows:

\begin{itemize}
    \item The MLP + NF model was trained using histograms of the subhalo properties from each simulation. This representation captures the full distribution shape of each property and allows identification of the most informative features for WDM mass inference.
    \item A second MLP + NF model was trained using the mean and standard deviation of the selected features for each simulation, enabling us to test whether the first two moments of the distribution are sufficient to constrain the WDM mass. Afterward, a symbolic regression model was used to derive an explicit mathematical expression linking the most informative subhalo property to the WDM mass.
    \item The GNN + NF model was trained using graphs built from individual halo information, where subhalo properties served as node features. To assess whether halo-level information provides additional predictive power beyond global simulation statistics, we included the mean and standard deviation used in the previous MLP + NF model as graph-level features. 
\end{itemize}

For the GNN model specifically, the datasets were constructed as described in Sec.~\ref{sec:graphdata}. We then associated the subhalos within each simulation with their respective host halo, restricting the analysis to halos that contain more than four subhalos. This threshold, corresponding to halos masses above $ \sim 1 \times 10^{11} h^{-1} M_{\odot}$, was selected as a compromise to ensure a sufficiently large sample for robust statistical analysis while excluding halos, with too few subhalos, that are unlikely to provide meaningful information.

For the training of the models, 720 simulations were used to construct the training set ($\thicksim 70\%$), 150 simulations for the validation set ($\thicksim 15\%$), and 154 simulations for the test set ($\thicksim 15\%$). The simulations were split in this manner to ensure that galaxies within the same simulation are grouped in the same dataset, as galaxies from the same simulation may share similar properties. This approach prevents potential information leakage between datasets, following the methodology outlined in \cite{2024ApJ...970..170L}. We employed the Adam Optimizer~\citep{2014arXiv1412.6980K} with L2 regularization on the weights. 

For hyperparameter tuning, we have used the \emph{Optuna} package~\citep{2019arXiv190710902A}, to find the combination of parameters that gave the best performance on the validation set. The tuned hyperparameters include the learning rate ($lr$), weight decay ($wd$), and, for the GNN model, the maximum linking distance between nodes ($r_{\rm{link}}$). We also optimized the number of graph layers or MLP layers (depending on the model) and the number of hidden units in each layer. For the normalizing flow component, we tuned the number of conditional layers and the number of bins. Tables~\ref{tab:mpl} and~\ref{tab:gnn} show the ranges adopted in each hyperparameter for the MLP + NF and GNN + NF models, respectively

\begin{table}[ht!]
\centering
\begin{tabular}{||c |c |c ||} 
 \hline
 \multicolumn{3}{|c|}{MLP + NF} \\
 \hline
  Hyperparameters & min & max  \\  
 \hline
 $lr$ & $1\times10^{-8} $ &  $1\times10^{-5}$\\ 
 \hline
 $wd$ & $1\times10^{-9}$ & $1\times10^{-6}$ \\
 \hline
 $N^{\circ}$ MLP layers & 1 & 5 \\
 \hline
 $N^{\circ}$ nodes & 32 & 256 \\ 
 \hline
 $N^{\circ}$ conditional layers & 1 & 10 \\
 \hline
 $N^{\circ}$ bins & 2 & 64 \\
 \hline
\end{tabular}
\caption{Range of the hyperparameters tuned for the MLP + NF model.}
\label{tab:mpl}
\end{table}

\begin{table}[ht!]
\centering
\begin{tabular}{||c |c |c ||} 
 \hline
 \multicolumn{3}{|c|}{GNN + NF} \\
 \hline
  Hyperparameters & min & max  \\  
 \hline
 $lr$ & $1\times10^{-8} $ &  $1\times10^{-5}$\\ 
 \hline
 $wd$ & $1\times10^{-9}$ & $1\times10^{-6}$ \\
 \hline
 $r_{\rm{link}}$ & $1\times10^{-8}$ & $1\times10^{-4}$ \\
 \hline
 $N^{\circ}$ graph layers & 1 & 5 \\
 \hline
 $N^{\circ}$ nodes & 64 & 512 \\ 
 \hline
 $N^{\circ}$ conditional layers & 1 & 15 \\
 \hline
 $N^{\circ}$ bins & 4 & 256 \\
 \hline
\end{tabular}
\caption{Range of the hyperparameters tuned for the GNN + NF model.}
\label{tab:gnn}
\end{table}

In the GNN + NF case, the batch size was set to 256, and at least 50 Optuna trials were performed. Each trial ran for 100 epochs with a unique combination of hyperparameters. For the MLP + NF case, the batch size was set to 32, and 100 Optuna trials were run.

Finally, the model with the lowest validation loss is selected and applied to the test set. This resulted in a posterior distribution $p(\vec{\theta}|\textbf{X})$, from which we drew samples and computed the  
median to compare with the target WDM mass (a single value per simulation). In addition, we computed the 16th, 50th, and 84th percentiles from the samples to estimate the 68\% credibility interval, offering a measure of uncertainty.

To evaluate the accuracy and precision of the models in predicting the WDM masses, we used the following metrics:

\begin{itemize}
    \item Mean relative error:
          \begin{equation}
              \epsilon = \frac{1}{N} \sum_{i}^{N} \frac{|y_{\rm true,i}-y_{\rm infer,i}|}{y_{\rm true,i}},
          \end{equation}
    \item Determination coefficient: 
          \begin{equation}
             R^{2} = 1 - \frac{\sum_{i}^{N}(y_{\rm true,i}-y_{\rm infer,i})^{2}}{\sum_{i}^{N}(y_{\rm true,i}-\bar{y}_{\rm true})^{2}}, 
          \end{equation}
    \item Root mean square error: 
          \begin{equation}
              RMSE = \sqrt{\frac{1}{N}\sum_{i}^{N}(y_{\rm true,i}-y_{\rm infer,i})^{2}},
          \end{equation}
    \item Expected coverage probability, which represents the average probability over different realizations that the true value of a parameter falls within the credibility interval~\citep{2023PMLR..20219256L}.
\end{itemize}

$N$ is the number of elements in the test set, $y_{\rm true,i}$ and $y_{\rm pred,i}$ denote the true and predicted values of the WDM mass, respectively. Then, $\bar{y}_{true}$ is the average of the true WDM mass values. 

A high $R^{2}$ (close to 1) and low \textit{RMSE} and $\epsilon$ indicate more accurate predictions. The expected coverage probability is used to evaluate how well the model’s credibility intervals capture the true parameter, thus assessing the reliability of the uncertainty estimation, as discussed in Appendix~\ref{apx1}.

\section{Results} 
\label{sec:results}

As described in Sec.~\ref{sec:methods}, we begin by identifying the galaxy property most sensitive to the WDM mass using the MLP + NF model, trained on histograms of individual galaxy properties. 
Next, we train another MLP + NF model using the mean and standard deviation of the subhalo properties across simulations. These two approaches enable us to evaluate the performance of WDM mass inference based solely on global information from the cosmological box, and to quantify the contribution of the full distribution shape and its first two moments.  To further interpret the results, we then apply symbolic regression to derive analytical expressions that relate the most predictive property to the WDM mass.

Finally, we compare the results of the MLP + NF model with those obtained from the GNN + NF model trained on detailed halo-level information. This allows us to determine whether incorporating information from individual halos, on top of global descriptors, leads to improved constraints.


\begin{figure*}
    \centering
    \includegraphics[width = 0.48\linewidth]{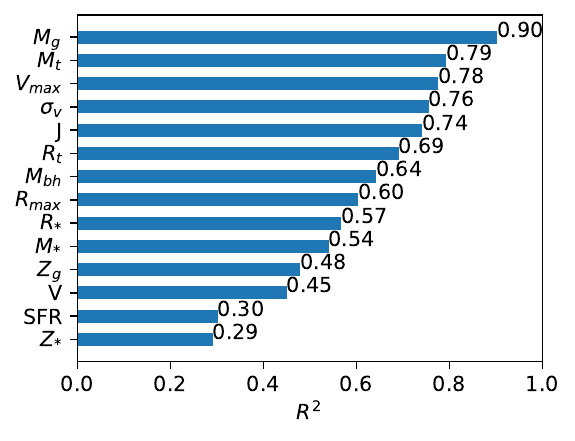}
    \includegraphics[width = 0.48\linewidth]{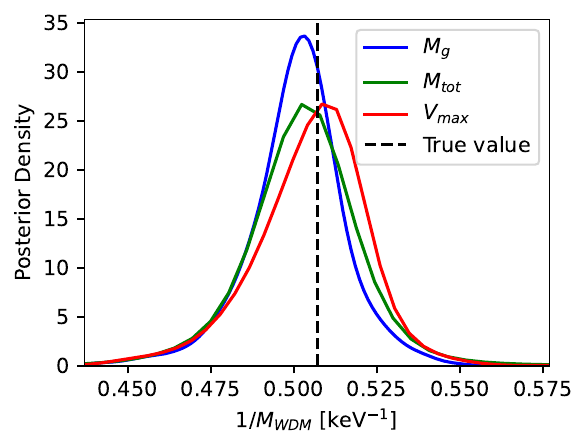}
    \caption{Feature importance analysis for predicting the warm dark matter (WDM) mass using histograms of various galaxy properties from all subhalos within a simulation box. The left panel displays the $R^2$ values for the inference. The right panel illustrates the shape of individual sample inferences from a selected simulation, highlighting the constraining power of $M_g$, $M_{\text{tot}}$, and $V_{\max}$. The dashed black line marks the true value of WDM mass for this chosen simulation.}
    \label{fig:feature_importance}
    
\end{figure*}

\subsection{WDM mass inference using global statistics}

\subsubsection{Galaxy Feature Importance}

To understand the relative contribution of different subhalo properties to constraining the WDM mass, we perform a feature importance analysis using histograms (with 14 bins) of each galaxy property as input. Specifically, we bin each property into 14 intervals using a customized binning strategy. When zero values are present in properties (e.g., $\mathrm{M}_g$, $\mathrm{M}_{*}$, $\mathrm{M}_{BH}$), they are separated into a dedicated bin, and then excluded from the remainder of the binning process. The range of remaining non-zero values is then determined between the global minimum and the 99.9th percentile to reduce the impact of outliers. This truncated range is then divided into evenly spaced bins. These histograms are then used as input features to a machine learning model that combines a multilayer perceptron (MLP) with a conditional normalizing flow, which learns the mapping between the distribution of individual galaxy properties in a simulation box and the corresponding WDM particle mass.

The left panel of Figure~\ref{fig:feature_importance} presents the coefficient of determination (\(R^2\)) for each subhalo property, quantifying the predictive power of each feature when used individually. The global gas mass (\(M_g\)) emerges as the most informative property, achieving an \(R^2\) of 0.90, followed by total mass (\(M_{\mathrm{tot}}\)), the maximum circular velocity (\(V_{\mathrm{max}}\)), and velocity dispersion ($\mathbf{\sigma_{\nu}}$). These features are strongly correlated with the underlying WDM mass and provide a robust statistical signal for the model to exploit. In contrast, properties such as the star formation rate (SFR), the modulus of the subhalo peculiar velocity (\textbf{V}), and gas metallicity (\(Z_g\)) yield substantially lower predictive power, indicating poor relevance or possible noise.

The right panel of Figure~\ref{fig:feature_importance} illustrates the inferred posterior distributions of WDM for a selected simulation using three of the top-performing features: \(M_g\), \(M_{\mathrm{tot}}\), and \(V_{\mathrm{max}}\). Among them, the distribution from \(M_g\) is the narrowest around the true value (indicated by the dashed black line), reflecting its strong constraining power and high precision. In contrast, the distributions from \(M_{\mathrm{tot}}\) and \(V_{\mathrm{max}}\) are noticeably broader, indicating greater uncertainty in their individual inferences and weaker constraints on the WDM mass. This analysis highlights that the most informative feature for WDM mass inference is the distribution of gas mass (\(M_g\)) within the simulation box. 

In Appendix~\ref{apx2}, we show the robustness of our results by restricting the analysis to subhalos with more than 30 and 50 dark matter particles, thus excluding objects with masses close to the resolution limit. Furthermore, we analyze the impact of the numerical fragmentation~\citep{2007MNRAS.380...93W} in our simulations to the WDM mass inference.

\subsubsection{Inference from First Two Moments}

To evaluate whether the WDM mass can be constrained using summary statistics rather than the full property distributions, we train an MLP + NF model using the mean and standard deviation (std) of the described subhalo properties. Properties with potential zero values (e.g., stellar mass or stellar metallicity in absence of starts) are handled by excluding zeros in the mean/std computation and including the count of zero-value subhalos as additional feature. Furthermore, the matter density $\Omega_{m}$ is included due to its influence on structure formation, and to investigate its specific role in the inference process, we also compare models trained with and without $\Omega_{m}$ as an input feature.

\begin{figure*}
    \includegraphics[width=1.\textwidth]{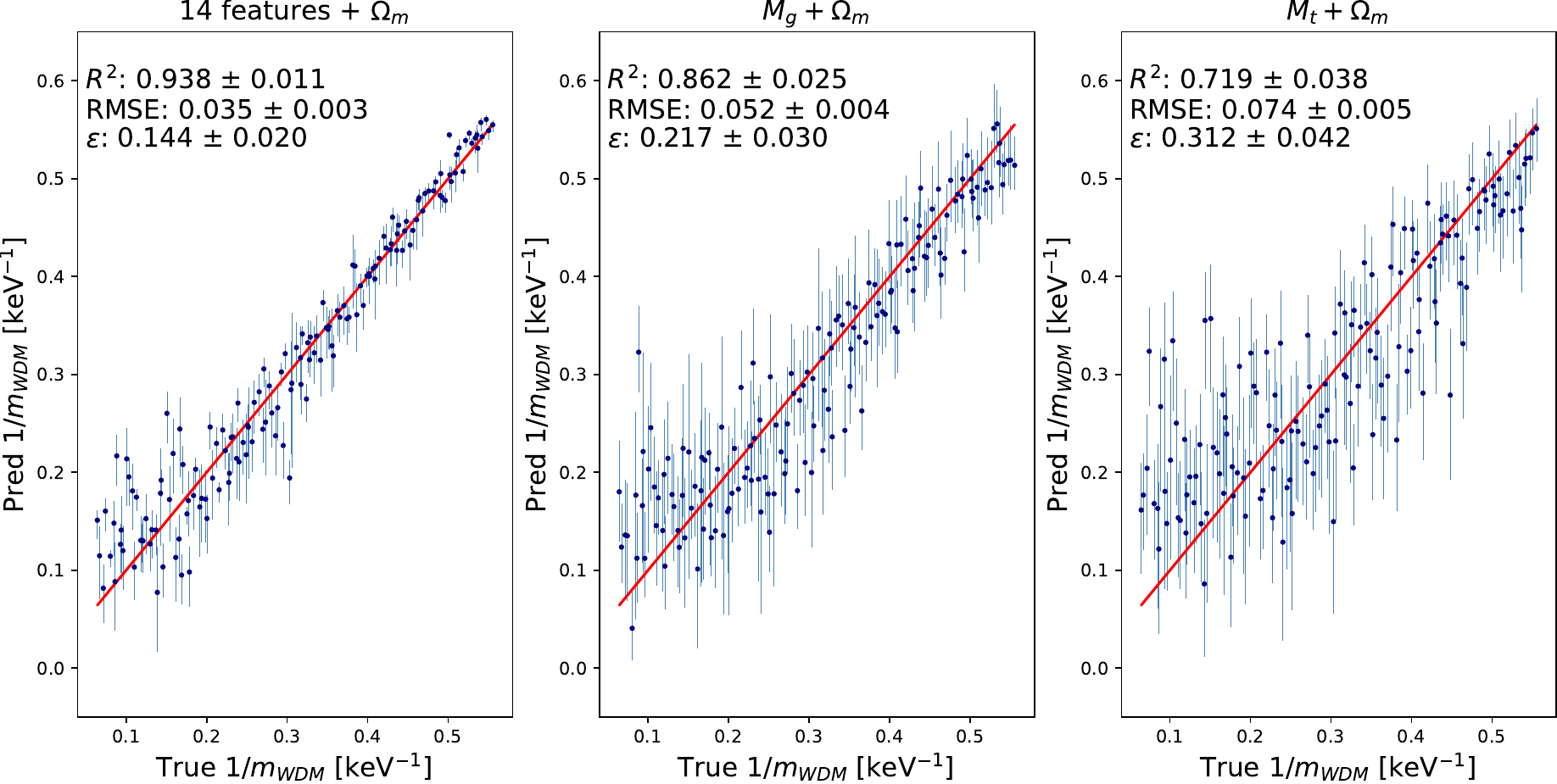}
    \caption{$1/m_{WDM}$ predictions (y-axis) using the MLP + NF model, compared to the true $1/m_{WDM}$ value of the WDM particle mass used in each simulation (x-axis). A one-to-one red line is shown to indicate where predictions match the true values. The left panel shows the results considering the 14 galaxy properties; the middle and right panels use only the gas mass $M_{g}$ and total mass $M_{t}$, respectively, as single features. Error bars correspond to the 68\% credible interval, computed as the [50th–16th, 84th–50th] percentiles of the posterior distribution.}
    \label{fig:MPL14}
\end{figure*}

Figure~\ref{fig:MPL14} shows predictions of the model. The left panel corresponds to the case where the input dataset consists of the mean and standard deviation of all 14 subhalo properties described in Sec.~\ref {sec:data} (i.e., a feature vector created by concatenating the mean and standard deviation of all features). The middle and right panels show the predictions when using only a single feature, gas mass and total mass, respectively, represented by their mean and standard deviation across subhalos in each simulation. These results correspond to the test set and were obtained using the best-performing model, selected with the lowest validation loss found by \emph{Optuna}.

As expected, using all features yields the best result with $R^{2}=0.94 \pm 0.01$. The uncertainty of the metric was estimated using a statistical method known as bootstrap resampling, in which the test set is repeatedly resampled by randomly selecting data points from the original test sample. By computing the $R^{2}$ score for each resampled set, we obtain an empirical distribution from which the standard deviation provides an estimate of the statistical uncertainty.   

When using only $M_{g}$, $R^{2}$ drops to $0.86 \pm 0.02$, and to $0.72 \pm 0.03$ when using only $M_{t}$. These results reinforce the conclusion that gas mass is the most informative single property for predicting the WDM mass, and that the mean and standard deviation of the distributions contain a significant fraction of the overall information embedded into the histograms. It's important to highlight that, for the subhalo gas mass, the feature vector is composed of the mean, std and the count of zero values. This is not the case for the subhalo total mass as it does not contain zero values.

\begin{table*}[!htbp]
    \centering
    \begin{tabular}{ll}
        \toprule
        \textbf{RMSE} & \textbf{Equation} \\
        0.074 & ${\rm keV}/m_{\rm WDM} = 0.31794^{\max(x_0, 1.317^{(x_0 - x_1)})}$ \\
        0.071 & ${\rm keV}/m_{\rm WDM} = 0.23954^{1.3354^{(x_0 - (x_1 + \cos(x_1)))}}$ \\
        0.068 & ${\rm keV}/m_{\rm WDM} = \cos(1.1321^{x_0})^{\max(1.6536 - x_1, 1.4427)}$ \\
        0.067 & ${\rm keV}/m_{\rm WDM} = \cos(0.89086^{(x_1 - x_0)})^{\max(1.8037, x_1 + x_0)}$ \\
        0.067 & ${\rm keV}/m_{\rm WDM} = \cos(1.1321^{\max(x_0, -1.3703)})^{\max(1.6536 - x_1, 1.4427)}$ \\
        0.066 & ${\rm keV}/m_{\rm WDM} = \max(\cos(0.88751^{(x_1 - x_0)})^{\max(1.8088, x_1 + x_0)}, 0.15375)$ \\
        \hline
        \hline
    \end{tabular}
    \caption{Fitted equations with symbolic regression using subhalo gas mass.}
    \label{tab:loss_equations}
\end{table*}

\subsubsection{Symbolic Relations}

To further investigate this relationship and gain interpretability, we perform a symbolic regression analysis using summary statistics derived from the gas mass distribution. Specifically, we use the mean gas mass and the total number of subhalos with \(M_g = 0\) in a simulation box, denoted as \(N_{\mathrm{ZeroGas}}\). The symbolic regression model identifies concise mathematical expressions that approximate the WDM mass based on these two variables. The resulting expressions are listed in Table~\ref{tab:loss_equations}, showing that even simple combinations of these gas-related quantities can capture meaningful trends and provide interpretable constraints on the underlying WDM particle mass. In the table, \(x_0\) and \(x_1\) denote the two normalized input variables used in the symbolic regression analysis: the number of subhalos with zero gas mass within the $(25h^{-1} \rm{Mpc})^{3}$ simulation box volume and the mean logarithmic gas mass across all subhalos, respectively. Their definitions are given as follows:

\begin{equation}
\begin{aligned}
x_0 &= \frac{N_{\mathrm{ZeroGas}} - 8957.3}{1418.3} , \\
x_1 &= \frac{\frac{1}{N}\sum_{i=1}^N\!log_{10}(1 + M_g^i/M_\odot) - 8.59}{0.16},
\end{aligned}
\end{equation}

where the $M_{g}^i$ represent the subhalo gas mass of the $i$-th subhalo within the simulation. The left column reports the root mean square error (RMSE) for each candidate expression, providing a direct measure of predictive accuracy comparable to the metrics used elsewhere in the paper.
Importantly, the best symbolic regression expression achieves an RMSE of 0.066, which is comparable to the errors obtained with our neural-network models. This demonstrates that the symbolic approach not only yields interpretable functional forms but also retains predictive power with more complex machine-learning methods.

\subsubsection{Importance of $\Omega_m$}

\begin{figure*}
    \includegraphics[width=1.\textwidth]{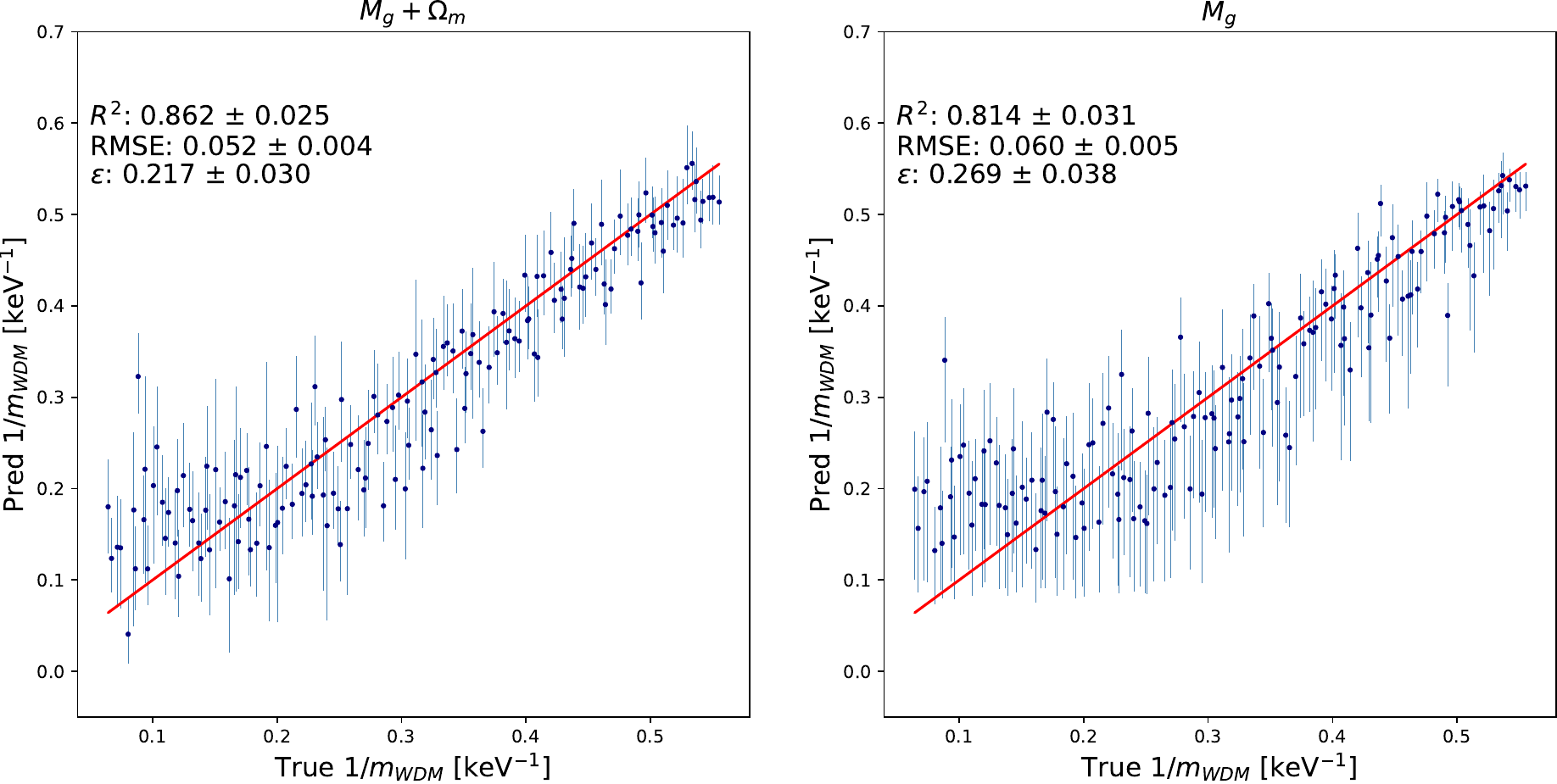}
    \caption{$1/m_{WDM}$ predictions (y-axis) using the MLP + NF model, compared to the true $1/m_{WDM}$ value of the WDM particle mass used to create each simulation (x-axis). A one-to-one red line is shown to indicate where predictions match the true values. On the left panel only the gas mass of the subhalo $M_{g}$ is used as a subhalo property, and the matter density of the simulation $\Omega_{m}$ is included, while on the right panel $\Omega_{m}$ is not included. Error bars correspond to the 68\% credible interval, computed as the [50th–16th, 84th–50th] percentiles of the posterior distribution.}
    \label{fig:MPLmtot}
\end{figure*}

In the middle panel of Figure~\ref{fig:MPL14}, we show the predictions obtained when the feature vector concatenates the mean and standard deviation of the subhalo gas mass, the count of zero values, and $\Omega_{m}$. To examine the impact of $\Omega_{m}$ on the inference of the WDM mass, we train a separate MLP + NF model using the same input features but excluding $\Omega_{m}$. Figure~\ref{fig:MPLmtot} compares the predictions for both cases: with $\Omega_{m}$ included (left panel) and excluded (right panel).

As expected, including $\Omega_{m}$ improves the model performance, as evidenced by a higher $R^{2}$ score. This is physically motivated by the fact that $\Omega_{m}$ governs the total matter content of the universe, and in particular, higher values of $\Omega_{m}$ correspond to higher dark matter densities. Since the dark matter distribution influences the gravitational potential wells in which galaxies form and evolve, it ultimately impacts its properties. Including $\Omega_{m}$ therefore allows the model to better capture variations in galaxy properties across different cosmological scenarios. Another way to see this is that $\Omega_m$ largely affects the abundance of subhalos, which is also affected by WDM mass. So being able to fix the value of $\Omega_m$ (or put a prior on it), is expected to improve the results. Besides, previous studies~\citep{2022ApJ...929..132V, Echeverri-Rojas_2023}, have explicitly demonstrated a strong correlation between $\Omega_{m}$ and the properties of individual galaxies. 

\subsection{WDM mass inference considering halo-level information}

We now compare the above results to those obtained from a GNN + NF model, which incorporates detailed halo-level information. The dataset consists of graphs, where each graph represents a dark matter halo and its nodes correspond to the subhalos within it. The node features are the 14 subhalo properties described in Sec.~\ref{sec:data}. Additionally, each graph includes global features, defined as the mean and standard deviation of the selected properties computed across the entire simulation in which the halo resides.

After hyperparameter tuning, the Optuna library provides the best model with the minimum loss found across several trials. We then evaluate the model on the test set, where, given the graphs from the simulations, the model outputs the conditional probability of the WDM particle mass given that specific graph. For every halo, we generate multiple realizations and compute the 16th, 50th, and 84th percentiles of the resulting posterior distribution. The median values (50th percentiles) are compared to the true WDM particle masses (shown on the x-axis) in the test set. These percentiles are also used to construct the 68\% credible interval.

\begin{figure}
    \includegraphics[width=1.\columnwidth]{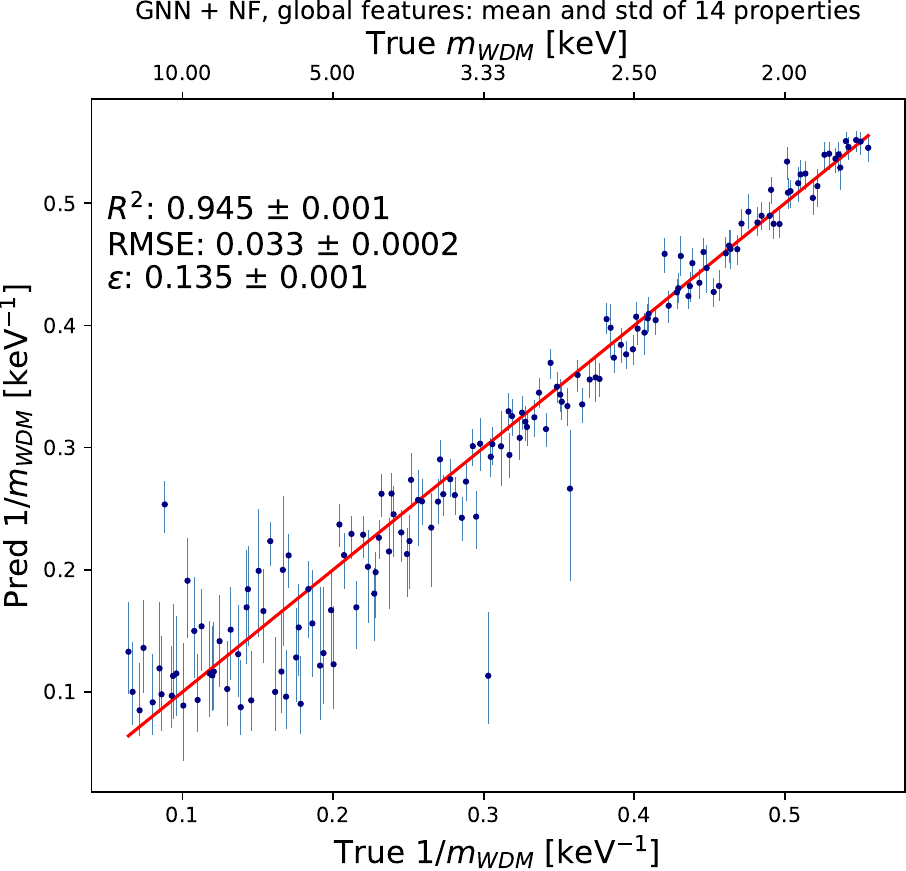}
    \caption{$1/m_{WDM}$ predictions (y-axis) using the GNN + NF model, compared to the true $1/m_{WDM}$ value of the WDM particle used to create each simulation (x-axis). A secondary axis at the top indicates the true $m_{WDM}$ value of the WDM. A one-to-one red line is shown to indicate where predictions match the true values. Node features consist of 14 subhalo properties, while global features correspond to the mean and standard deviation of these properties across all subhalos in each simulation. Error bars correspond to the 68\% credible interval, computed as the [50th–16th, 84th–50th] percentiles of the posterior distribution.}
    \label{fig:GNN14}
\end{figure}

Figure~\ref{fig:GNN14} displays the model predictions (one halo per simulation is drawn) when using the 14 galaxy properties as node features and, as graph-level features, the mean and standard deviation of these properties computed across all subhalos in the same simulation from which each individual halo (graph) is drawn. These results are highly accurate across the entire mass range, specifically up to 4 keV, with small uncertainties. However, beyond 6 keV, the uncertainties increase. This behavior is expected for a model trained to predict the WDM mass, as higher masses approach the cold dark-matter (CDM) regime, making it more challenging to distinguish them. 

One contributing factor is the distribution of simulations across WDM masses. As shown in Figure~\ref{fig:wdm_range}, fewer simulations exist in the high mass CDM regime compared to the WDM regime. This imbalance introduces an implicit prior that favors WDM-like masses, which can influence the predictions and the associated uncertainties.

At the same time, a few outliers are present. This behavior is expected given the variety of cosmological and astrophysical parameters in our simulations, together with the intrinsic scatter in galaxy properties. For example, halos with high $\Omega_m$ and weak feedback, or those whose evolution is strongly affected by environmental conditions (e.g. highly isolated regions or very dense environments), can naturally sit in the tail of the distribution. Consequently, their individual inference may appear “biased” but such cases are less representative of the global trend. Importantly, the model remains unbiased at the population level.

This GNN + NF model achieves an $R^{2}=0.945 \pm 0.001$, which is nearly identical to the result from the MLP + NF model using only global statistics ($R^{2} = 0.94 \pm 0.01$; see left panel of Figure~\ref{fig:MPL14}). This indicates that most of the information is coming from the global features, while the subhalo properties within individual halos contribute only residual information.

The smaller uncertainty in the GNN + NF model ($\pm 0.001$), compared to the MLP + NF model ($\pm 0.01$), is due to the larger size of the test set: the GNN + NF model produces one prediction per halo, resulting in many more data points per simulation (as each simulation contains several halos). In contrast, the MLP + NF relies on a single global summary per simulation, leading to greater variability in bootstrap resampling.


%

In order to build a model with a GNN, we take into account the edges between galaxies, where the nodes receive information from their neighborhoods. The maximum distance between two galaxies required to be connected by an edge is a hyperparameter of the model. We experimented with smaller and larger distance values, resulting in connected graphs or very sparse graphs (with many nodes having no connections). 

\begin{figure}
    \includegraphics[width=1.\columnwidth]{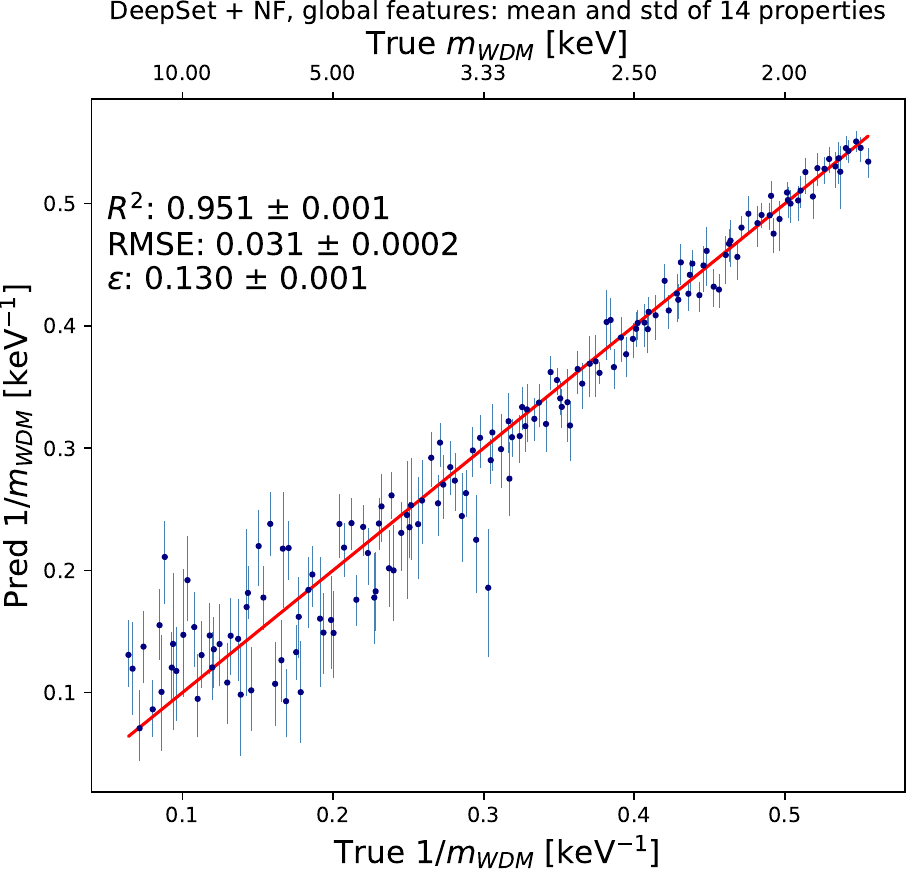}
    \caption{$1/m_{WDM}$ predictions (y-axis) using the DeepSet + NF model, compared to the true $1/m_{WDM}$ value of the WDM particle used to create each simulation (x-axis). A secondary axis at the top indicates the true $m_{WDM}$ value of the WDM. A one-to-one red line is shown to indicate where predictions match the true values. Node features consist of 14 subhalo properties, while global features correspond to the mean and standard deviation of these properties across all subhalos in each simulation. Error bars correspond to the 68\% credible interval, computed as the [50th–16th, 84th–50th] percentiles of the posterior distribution.}
    \label{fig:deep14}
\end{figure}

Figure~\ref{fig:deep14} shows the predictions (y-axis) using a DeepSet model (explained in Sec.~\ref{sec:GNN}), which does not incorporate edges information, but the same information for the nodes and global features as the GNN model. The predictions agree quite well among the entire WDM mass range, but even better up to 4 keV, similar to the previous result in Figure~\ref{fig:GNN14}, and the determination coefficient $R^{2}$ is equal to $0.95 \pm 0.001$ which is closer to 1 as well.

This result suggests that the relevant information that the neural network needs for an accurate WDM mass prediction and posterior estimation does not primarily come from the clustering of the subhalos, as their positions were not included among the node features. To reinforce this point, we note that the best hyperparameter found by Optuna in the GNN model, is with a very small $r_{link}$ equal to $7 \times 10^{-6}$ (i.e. even the GNN preferred to discard spatial information).

\begin{figure*}
    \centering
   \includegraphics[width=1.\textwidth]{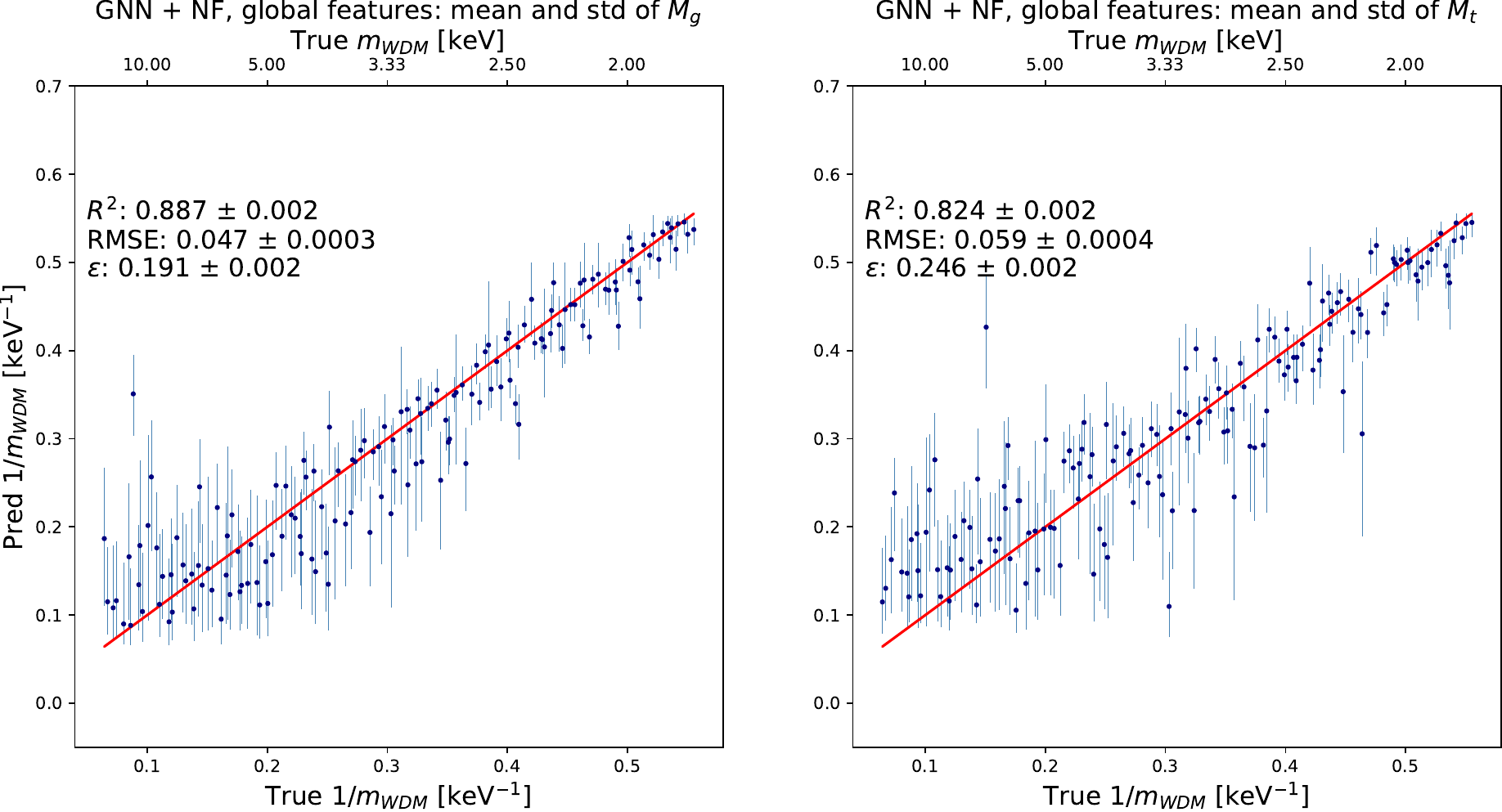}
    \caption{Left panel: GNN + NF model using the 14 nodes features and only the mean and standard deviation of the subhalo gas mass as global features. Right panel: GNN + NF model using the 14 nodes features and only the mean and standard deviation of the subhalo total mass as global features. In both panels the error bars correspond to the 68\% credible interval, computed as the [50th–16th, 84th–50th] percentiles of the posterior distribution.}
    \label{fig:mtot_mbh}
\end{figure*}

We train now the GNN + NF model, presented in Figure~\ref{fig:mtot_mbh}, incorporating the 14 galaxy properties as node features and, as global features, the mean and standard deviation of the subhalo gas mass (left panel) or the subhalo total mass (right panel). We compare these results to the middle and right panels of Figure~\ref{fig:MPL14}, which show the performance of the MLP + NF models trained only on the corresponding global features. This comparison allows us to assess the improvement obtained by incorporating detailed halo structures in the GNN-based approach.

In both panels, the predictions improve when the node features are included, compared to using only the mean and standard deviation of $M_{g}$ or $M_{t}$. Specifically, for the total mass $M_{t}$, the model achieves a $R^{2} = 0.82 \pm 0.002$, an improvement over the $R^{2}=0.72 \pm 0.03$ reported in the right panel of Figure~\ref{fig:MPL14}, with no overlaps in the uncertainty ranges. For the gas mass $M_{g}$ case, the $R^{2}$ increases to $0.89 \pm 0.002$ compared to $R^{2} = 0.86 \pm 0.02$ in the left panel of Figure~\ref{fig:MPL14}. Although the uncertainty intervals are close, the improvement remains consistent with the addition of node features.

A similar improvement is observed in the mean relative error, $\epsilon$. For the total mass $M_{t}$, the GNN model achieves an accuracy of $24\%$, while the MLP model reaches $31\%$. For the gas mass $M_{g}$, the GNN model attains $19\%$ of accuracy, whereas the MLP model achieves $21\%$.   
This demonstrates that adding node-level subhalo properties enhances the model's ability to extract meaningful information and achieve better predictions, than only using the global information of the gas mass or subhalo total mass. 

Thus, the inclusion of information from individual halos leads to improved predictions of the WDM particle mass compared to using only the global features of the simulations. This indicates that small-scale structures hold residual information for constraining WDM models, although the primary source of information are the global properties of the cosmological box. Furthermore, the approach enables the posterior distribution of the WDM mass to be inferred for each individual halo.

So far we were analyzing the accuracy of the results, to investigate the estimation of the uncertainties in Appendix~\ref{apx1} we present the expected coverage probability of all the NF models used.

\section{Conclusions}
\label{sec:conclusions}


In previous work~\citep{2024ApJ...970..170L}, it was found that the changes induced by WDM do not significantly affect simulated galaxies properties. It was demonstrated that the inference of WDM mass using individual galaxy properties alone is not feasible.

In this work, we revisit this challenge by first exploring whether summary statistics of galaxy populations across cosmological boxes can be used to infer the WDM mass. Using a set of 1024 high-resolution hydrodynamic simulations from the DREAMS project~\citep{2025ApJ...982...68R}, which were specifically designed to probe the imprints of dark matter models beyond cold dark matter (CDM), we train a model composed of a Multilayer Perceptron (MLP) and Normalizing Flows (NF). Each simulation is summarized using descriptors such as the mean, standard deviation, or histogram of 14 key galaxy properties. We find that the best predictive performance is achieved when using the mean and standard deviation of these 14 features. The resulting model achieves an $R^2 = 0.94 \pm 0.01$, demonstrating that global statistical properties of the galaxy population, in particular the first two moments of the properties distribution, encode sufficient information to accurately recover the WDM mass (see Figure~\ref{fig:MPL14}).

To further understand the relative importance of each galaxy property, we train a series of MLP + NF models, each using the histogram of a single feature. We find that subhalo gas mass distribution provides the strongest constraint on the WDM mass, exceeding even the total mass as it can be noticed in Figure~\ref{fig:feature_importance}. This result is further examined in Appendix~\ref{apx2}, where we restrict the analysis to subhalos with more than 30 and more than 50 dark matter particles, in order to exclude poorly resolved subhalos. We find that the conclusion remains robust: the subhalo gas mass function is consistently more informative than the total mass distribution for predicting the WDM mass.  

This robustness across different DM particle number thresholds demonstrates that our results are not significantly affected by spurious halos arising from numerical fragmentation, as further discussed in Appendix~\ref{apx2}.

The importance of the subhalo gas mass is also consistent with the middle and right panels of Figure~\ref{fig:MPL14}, where the model that uses the mean and standard deviation of the gas mass outperforms the model using the total subhalo mass. In Table~\ref{tab:loss_equations} we present mathematical expressions fitted with symbolic regression, that approximate the WDM mass given the subhalo gas mass information, specifically, the mean gas mass and the number of subhalos with $M_{g} = 0$. 

We then investigate whether the performance of the global statistics can be improved by incorporating more localized, halo-level information. Specifically, we represent each dark matter halo as a graph and use a Graph Neural Network (GNN) + NF model to infer the WDM mass based on individual halos. The graph includes node features (the 14 properties of subhalos), global features (the mean and standard deviation of those properties across all subhalos in the simulation where the halo resides) and edges connecting the subhalos (although we later test the effect of removing edge features). The model outputs the posterior distribution of the WDM particle mass given each halo-level graph.

We find that using the 14 subhalo properties as node features does not significantly improve the predictions when the global simulation statistics are already included as graph-level features. This suggests that most of the predictive information is already captured by the global descriptors, and little is gained from additional halo-level details (see Figure~\ref{fig:GNN14}). When the global input is limited to a single galaxy property (e.g., subhalo gas mass or total mass), the node features in the halos become more informative, leading to slightly improved performance. This comparison demonstrates that relevant WDM signatures are primarily encoded in large scale galaxy population statistics rather than in detailed halo-level features (see Figures~\ref{fig:MPL14} and \ref{fig:mtot_mbh}).

In particular, Figure~\ref{fig:deep14} demonstrates that no edge features are needed to constrain WDM mass accurately, as the results are very similar to those obtained with the GNN model. 

These findings suggest that the gas mass distribution alone is sufficient to estimate the WDM particle mass. In particular, the number of subhalos with zero gas mass emerges as a highly informative variable. It plays a key role in the symbolic regression expressions and significantly improves the constraint on the WDM mass when combined with the mean and standard deviation of the gas mass distribution. Therefore, the number of subhalos with no gas becomes a key discriminant in distinguishing between WDM and CDM models in our analysis.

In the context of structure formation in WDM cosmologies, the suppression of small scale power delays the collapse of low mass subhalos~\citep{2014JCAP...08..007S, 2012MNRAS.421.2384M, 2014MNRAS.437..293H}. A possible explanation for the importance of gas mass absence is that these subhalos tend to accrete and retain less gas due to their shallower potential wells and less efficient cooling mechanism. This may led to a larger fraction of subhalos without gas in WDM scenarios compared to CDM. 

Since atomic hydrogen ($\mathrm{HI}$) traces the distribution and kinematics of the neutral gas in galaxies~\citep{2025arXiv250303818M, 2025arXiv250308999X}, $\mathrm{HI}$ observations could serve as a promising avenue to further constrain the WDM mass using real data. 

In a previous work~\citep{2014JCAP...08..007S}, it was shown that WDM models affect the formation of minihalos ($M < 10^{7} M_{\odot}$, without star formation) and induce 21 cm line fluctuations at high redshifts ($z \gtrsim 5$). This demonstrates the potential of using 21 cm observations to constrain the WDM mass at early cosmic times. In contrast, our results highlight a complementary avenue at $z = 0$, where the absence of gas in subhalos can be used as a tracer of WDM effects on galaxy formation.

In addition, as a future work, we plan to relax the current threshold of requiring halos to contain more than four subhalos, to explore the WDM mass inference considering a wider mass range of halos.   


\section{Acknowledgements}
\begin{acknowledgments}
 This work was supported by the National Science Foundation under Cooperative Agreement 2421782 and the Simons Foundation award MPS-AI-00010515. B.C. acknowledges a doctoral fellowship by CONICET. The work of FVN is supported by the Simons Foundation. 
\end{acknowledgments}

\appendix

\section{Coverage test}
\label{apx1}

We have trained MLP + NF and GNN + NF models that output the posterior distribution $p(m_{WDM}|\vec{X})$, where $\vec{X}$ denotes the corresponding input statistics. To assess the calibration of the inferred posteriors, we generate samples from the learned distributions and compute the coverage probability by comparing the standard deviation of the predictions with the expected credible intervals.

\begin{figure*}
    \centering
   \includegraphics[width=1.\textwidth]{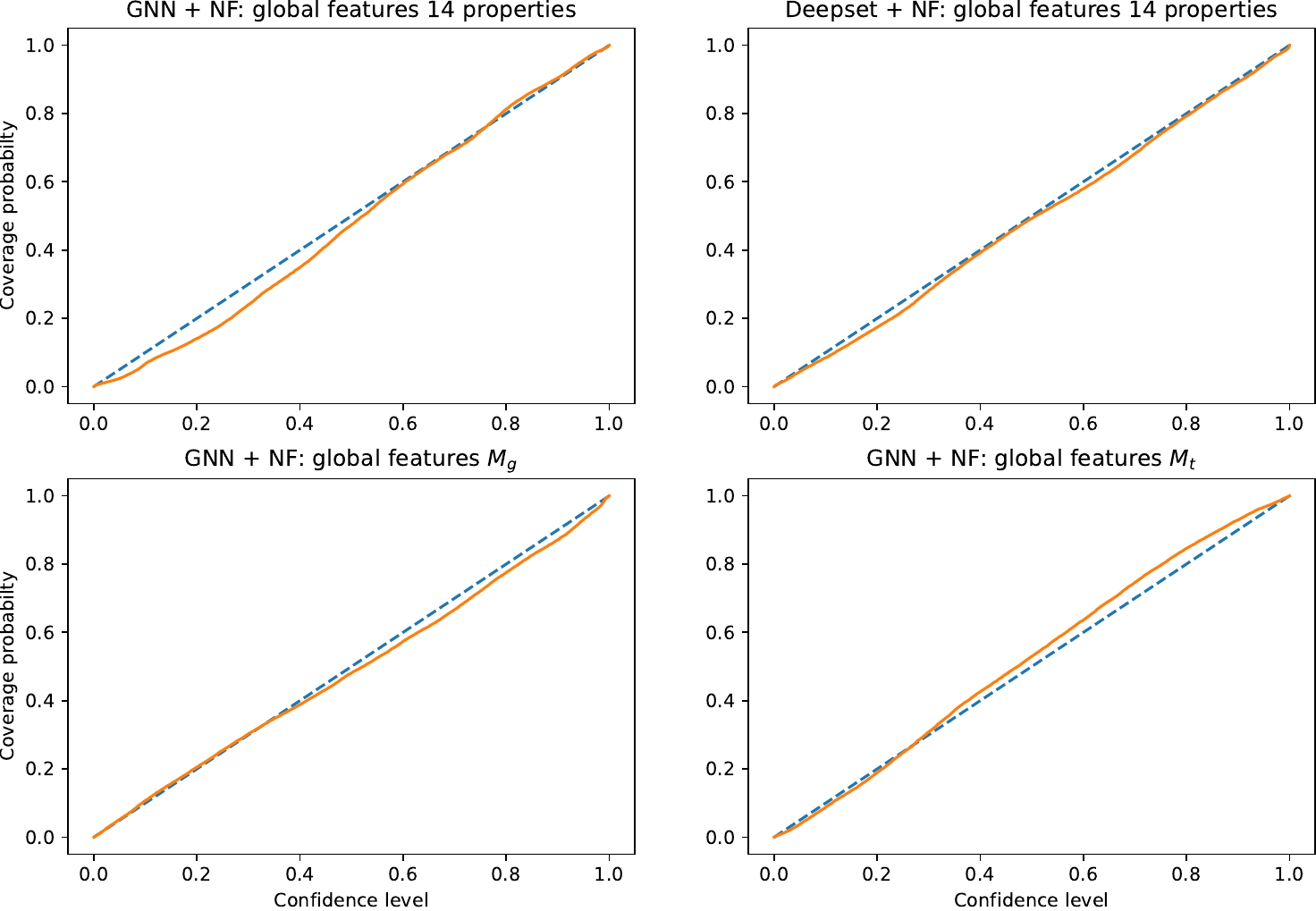}
    \caption{In the x-axis it is presented the confidence level, while in the y-axis the coverage probability. The top panels present the results when the mean and standard deviation of the 14 properties are used as global features (in the left panel the GNN + NF model, in the right panel the Deepset + NF model). On the bottom panels are presented the results when the mean and standard deviation of the $M_{g}$ (left panel) and $M_{t}$ (right panel) are used as global features.}
    \label{fig:cov}
\end{figure*}

Figure~\ref{fig:cov} shows the expected coverage probability of NF models, when the GNN is trained to generate the latent space. At the top it is presented the coverage probability when considering as global features the mean and standard deviation of the 14 properties. On the bottom the coverage probability considering as global features only the gas mass $M_{g}$ (left panel) or the total mass $M_{t}$ (right panel). 

If the model estimates the uncertainties accurately, the coverage probability should match the specified credibility level. However, an "S"-shaped pattern in the coverage plot indicates an overestimation or underestimation of the standard deviation. Using the \textsc{tarp} method, we generate 1000 samples from the posterior distribution and compute the coverage probability. As shown in Figure~\ref{fig:cov}, the uncertainties are sufficiently well calibrated, with slightly overestimation or underestimation depending of the case. 

\begin{figure*}
    \centering
   \includegraphics[width=1.\textwidth]{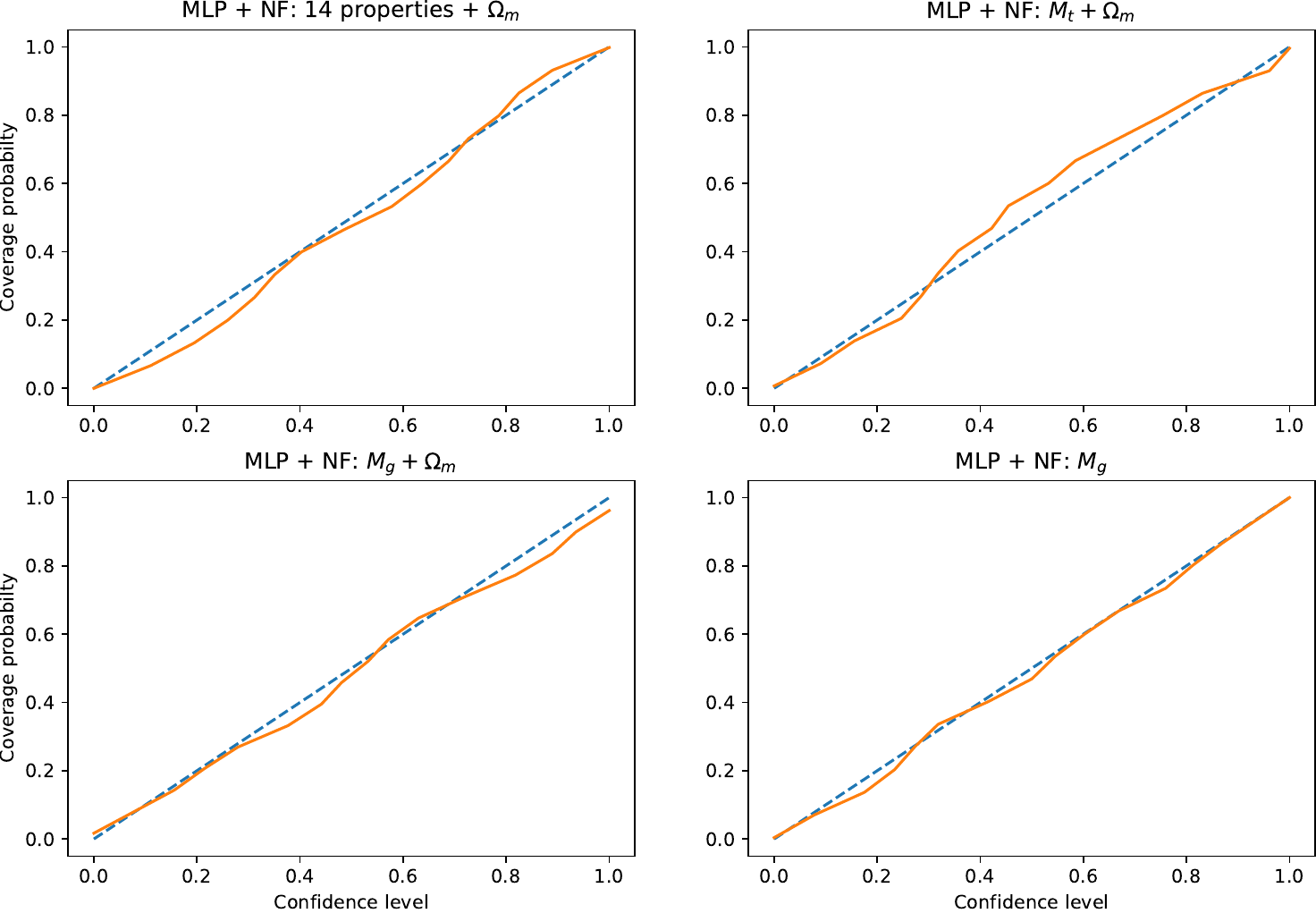}
    \caption{In the x-axis it is presented the confidence level, while in the y-axis the coverage probability for the models trained with MLP + NF. The top left panel present the results when the mean and standard deviation of the 14 properties are used, while the top right present the case using only the mean and standard deviation of the $M_{t}$. On the bottom panels are presented the results using the mean and standard deviation of the $M_{g}$ with $\Omega_{m}$ included (left panel) and not included (right panel).}
    \label{fig:covmpl}
\end{figure*}

On the other hand, Figure~\ref{fig:covmpl} presents the expected coverage probability of the NF models when the MLP is trained to create the latent space using global information from the simulations (specifically, the mean and standard deviation of the selected features). The top-left panel shows the case where all 14 properties are considered, while the top-right panel corresponds to the scenario where only the subhalo total mass is used. The bottom panels present the results when the subhalo gas mass is considered, where the bottom-right panel illustrates the case excluding $\Omega_{m}$.

\section{Effect of excluding poorly resolved subhalos}
\label{apx2}


In the main analysis of this work, all subhalos were initially included. To test the robustness of our conclusions, we repeated the feature importance analysis but restricted the sample to subhalos containing more than 30 DM particles first, and subsequently to those with more than 50 DM particles. 

\begin{figure*}
    \centering
   \includegraphics[width=0.9\textwidth]{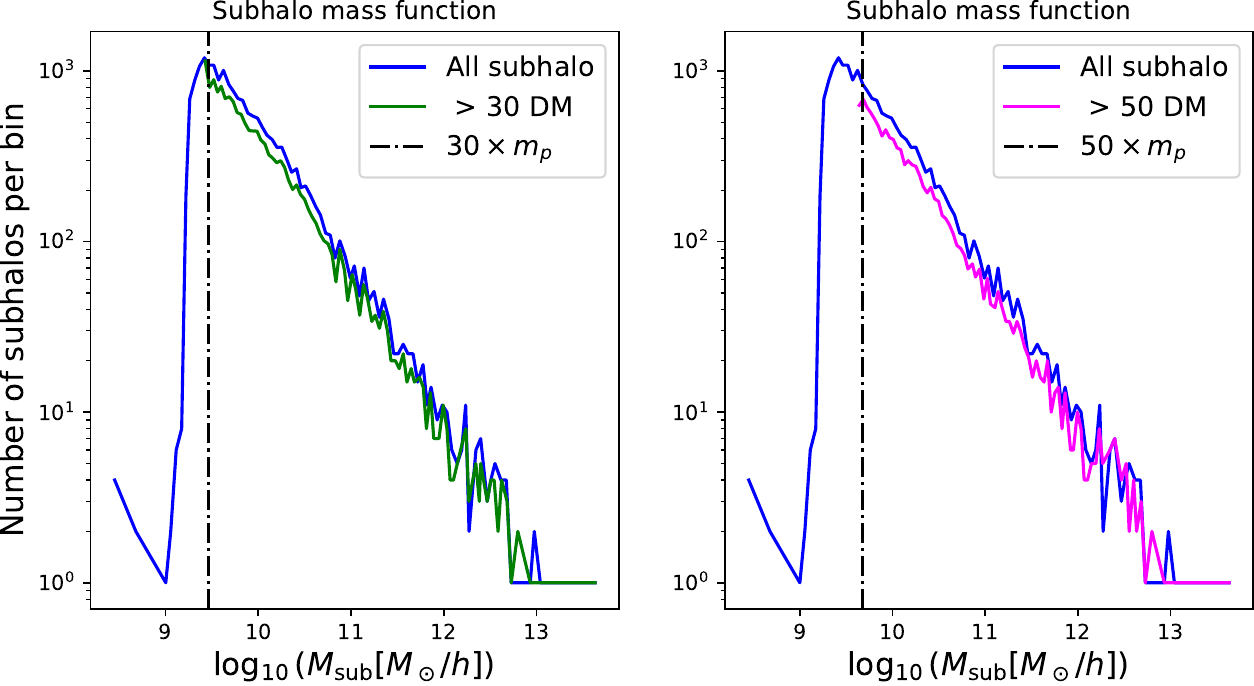}
    \caption{Left panel: SHMF considering all the subhalos in the simulation (blue curve) and SHMF considering subhalos with more than 30 DM particles (green curve). Right panel: SHMF considering all the subhalos in the simulation (blue curve) and SHMF considering subhalos with more than 50 DM particles (pink curve). The dashed black lines represent the masses $30 \times m_{p}$ and $50 \times m_{p}$, where $m_{p}$ is the DM mass resolution.}
    \label{fig:SHMF}
\end{figure*}

The DM mass resolution of the simulations used in this work is $7.81 \times (\Omega_{\rm m}/0.302) \times 10^{7}~h^{-1} M_{\odot}$. In Figure~\ref{fig:SHMF}, we present the subhalo mass function (SHMF) considering all the subhalos present in one simulation (blue curve in both panels). In the left panel, the green curve shows the SHMF when only subhalos with more than 30 dark matter particles are considered, while in the right panel, the pink curve corresponds to subhalos with more than 50 particles. It can be observed that subhalos composed of fewer than $\sim 30$ DM particles may suffer from numerical instabilities and are generally considered to be poorly resolved. Although the exclusion of low-mass subhalos reduces the number of objects at the low-mass end, the overall shape of the mass function is preserved above the resolution threshold. 

\begin{figure}
    \centering
    \includegraphics[width=0.45\linewidth, trim=0 0 0 0, clip]{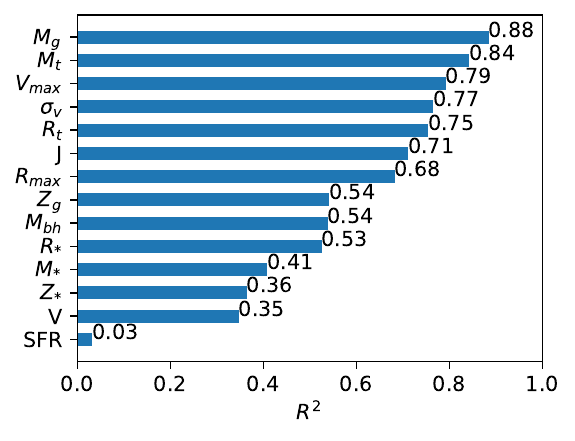}
    \includegraphics[width=0.47\linewidth]{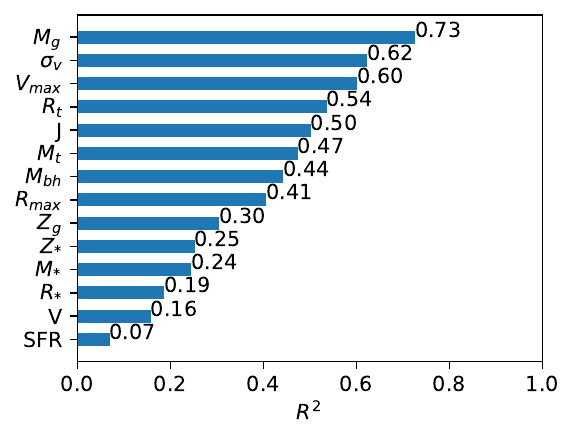}
    \caption{Feature importance analysis for predicting the warm dark matter (WDM) mass using histograms of various galaxy properties from selected subhalos within a simulation box. The left panel displays the $R^2$ values for the inference using subhalos with more than 30 DM particles while the right panel displays using subhalos with more than 50 DM particles. }
    \label{fig:enter-label}
\end{figure}

The left panel of Figure~\ref{fig:enter-label} shows the $R^{2}$ value for models trained using the histogram of each individual property, considering suhalos with more than 30 DM particles. The right panel presents the same analysis but using a more conservative selection of subhalos with more than 50 DM particles. Reducing the number of subhalos leads to a decrease in predictive performance due to the smaller dataset size. However, the gas mass consistently remains the most informative galaxy property for constraining the WDM mass.

Another issue to address is the numerical fragmentation~\citep{2007MNRAS.380...93W} inherent to simulations with low power on small scales, as in the WDM case. This effect arises from discrete sampling of the underlying density distribution, where the artificial power on smaller scales can grow and lead to the formation of spurious halos. According to~\citep{2007MNRAS.380...93W}, the typical scale of these artificial halos can be estimated as $M_{\rm{lim}} \approx 10.1 \bar{\rho} d k_{\rm{peak}}^{-2}$, where $\bar{\rho}$ is the mean density of the universe, $d$ is the mean interparticle separation, and $k_{\rm peak}$ is the wavenumber at which the dimensionless linear power spectrum reaches its maximum.

Applying this formula to our warmest model with $\Omega_m = 0.5$ and $m_{WDM} = 1.8 \rm{keV}$, the estimated $M_{\rm{lim}}$ is $\sim 8 \times 10^{8} h^{-1} M_{\odot}$. The resolution of the simulation with $\Omega_m = 0.5$ and $m_{WDM} = 1.8 \rm{keV}$ is $1.3 \times 10^{8} h^{-1} M_{\odot}$. Then, the spurious halos expected from numerical fragmentation would be composed of only $\sim 6$ DM particles. For larger WDM particle masses, the impact of numerical fragmentation is progressively less pronounced.

In Figure~\ref{fig:enter-label} we show the $R^{2}$ values of the WDM mass inference when selecting only subhalos with more than 30 and 50 DM particles. This confirms that the predictive power of our models originates from physically resolved halos rather than from unresolved or spurious ones. Therefore, our results are robust against the effects of numerical fragmentation. Moreover, we find that the subhalo gas mass consistently remains the most predictive feature, independently of the particle number threshold applied.

In a previous work~\citep{2024MNRAS.527..739R}, based on simulations similar to ours, it was shown that numerical fragmentation starts to counteract the genuine WDM suppression for particle masses just below $1.8 \rm{keV}$. An alternative strategy to mitigate the impact of spurious halos was proposed in~\cite{2014MNRAS.439..300L}, where halos were identified and filtered based on their characteristic shapes.

\bibliography{bibliography_wdm}{}
\bibliographystyle{aasjournal}

\end{document}